\documentclass[fleqn,usenatbib]{mnras}

\usepackage{newtxtext,newtxmath}

\usepackage[T1]{fontenc}

\DeclareRobustCommand{\VAN}[3]{#2}
\let\VANthebibliography\thebibliography
\def\thebibliography{\DeclareRobustCommand{\VAN}[3]{##3}\VANthebibliography}

\usepackage{graphicx}	
\usepackage{amsmath}	
\usepackage[draft,inline,nomargin,index]{fixme}
\fxsetup{theme=color,mode=multiuser}
\FXRegisterAuthor{af}{aaf}{\color{purple}Felipe}
\usepackage{graphics}
\usepackage{arydshln}
\usepackage{soul}

\newcommand{\abund}[2]{\ensuremath{[\mathrm{#1}/\mathrm{#2}]}}
\newcommand{\afe}{\abund{\alpha}{Fe}}

\newcommand{\metal}{\abund{Fe}{H}}
\newcommand{\teff}{\ensuremath{T_\mathrm{eff}}}
\newcommand{\mass}{\ensuremath{m_\mathrm{ini}}}
\newcommand{\logg}{\ensuremath{\log\,g}}
\newcommand{\Rapo}{\ensuremath{R_\mathrm{apo}}}
\newcommand{\Rperi}{\ensuremath{R_\mathrm{peri}}}
\newcommand{\zmax}{\ensuremath{z_\mathrm{max}}}
\newcommand{\ecc}{\ensuremath{\mathrm{ecc}}}

\newcommand{\kms}{\ensuremath{\mathrm{km}\,\mathrm{s}^{-1}}}

\title[Properties of Metal-Poor Stars in S-PLUS]{Chemodynamical Properties and Ages of Metal-Poor Stars in S-PLUS}

\author[Almeida-Fernandes et al.]{
F. {Almeida-Fernandes}\thanks{E-mail: felipe.almeida.fernandes@usp.br, felipefer42@gmail.com},$^{1, 2}$
V. M. {Placco},$^{2}$
H. J. {Rocha-Pinto},$^{3}$
M. {Borges Fernandes},$^{4}$ 
G. {Limberg},$^{1,5,6}$ 
\newauthor{
L. {Beraldo e Silva},$^{7}$
J. A. S. {Amarante}\thanks{Visiting Fellow at UCLan},$^{8,9}$
H. D. {Perottoni},$^{1,8}$
R. Overzier,$^{1,4}$
W. Schoenell,$^{10}$}
\newauthor{
T. Ribeiro,$^{11}$
A. Kanaan,$^{12}$
C. Mendes de Oliveira$^{1}$
}
\\
$^{1}$Universidade de S\~ao Paulo, Instituto de Astronomia, Geof\'isica e Ci\^encias Atmosf\'ericas, Departamento de Astronomia, SP 05508-090, S\~ao Paulo, Brazil\\
$^{2}$NSF’s NOIRLab, 950 N. Cherry Ave., Tucson, AZ 85719, USA\\
$^{3}$Observat\'orio do Valongo, Universidade Federal do Rio de Janeiro – UFRJ, Ladeira Pedro Ant\^onio 43, 20080-090 Rio de Janeiro, RJ, Brazil \\
$^{4}$Observat\'orio Nacional, Rua General Jos\'e Cristino 77, CEP: 20921-400, S\~ao Crist\'ov\~ao, Rio de Janeiro, Brazil \\
$^{5}$Department of Astronomy \& Astrophysics, University of Chicago, 5640 S. Ellis Avenue, Chicago, IL 60637, USA \\
$^{6}$Kavli Institute for Cosmological Physics, University of Chicago, Chicago, IL 60637, USA \\
$^{7}$ Department of Astronomy, University of Michigan, 1085 S. University Ave., Ann Arbor, MI 48109, USA \\
$^{8}$ Institut de Ciencies del Cosmos (ICCUB), Universitat de Barcelona (IEEC-UB), Martí i Franquès 1, E08028 Barcelona, Spain \\
$^{9}$ Jeremiah Horrocks Institute, University of Central Lancashire, Preston, PR1 2HE, UK \\
$^{10}$GMTO Corporation 465 N. Halstead Street, Suite 250 Pasadena, CA 91107, USA \\
$^{11}$Rubin Observatory Project Office, 950 N. Cherry Ave., Tucson, AZ 85719, USA \\
$^{12}$Departamento de F\'isica, Universidade Federal de Santa Catarina, Florian\'opolis, SC 88040-900, Brazil
}

\date{Accepted 15-May-2023. Received 22-Feb-2023}

\pubyear{2023}

\begin{document}
\label{firstpage}
\pagerange{\pageref{firstpage}--\pageref{lastpage}}
\maketitle

\begin{abstract}
Metal-poor stars are key to our understanding of the early stages of chemical evolution in the Universe. New multi-filter surveys, such as the Southern Photometric Local Universe Survey (S-PLUS), are greatly advancing our ability to select low-metallicity stars. In this work, we analyse the chemodynamical properties and ages of 522 metal-poor candidates selected from the S-PLUS data release 3. About 92\% of these stars were confirmed to be metal-poor ([Fe/H] $\leq -1$) based on previous medium-resolution spectroscopy. We calculated the dynamical properties of a subsample containing 241 stars, using the astrometry from Gaia Data Release 3. Stellar ages are estimated by a Bayesian isochronal method formalized in this work. We analyse the metallicity distribution of these metal-poor candidates separated into different subgroups of total velocity, dynamical properties, and ages. Our results are used to propose further restrictions to optimize the selection of metal-poor candidates in S-PLUS. The proposed astrometric selection ($\mathrm{parallax}>0.85$ mas) is the one that returns the highest fraction of extremely metal-poor stars (16.3\% have $\metal \leq -3$); the combined selection provides the highest fraction of very metal-poor stars (91.0\%  have $\metal \leq -2$), whereas the dynamical selection (eccentricity > 0.35 and diskness < 0.75) is better for targetting metal-poor (99.5\% have $\metal \leq -1$). Using only S-PLUS photometric selections, it is possible to achieve selection fractions of 15.6\%, 88.5\% and 98.3\% for metallicities below $-$3, $-$2 and $-$1, respectively. We also show that it is possible to use S-PLUS to target metal-poor stars in halo substructures such as Gaia-Sausage/Enceladus, Sequoia, Thamnos and the Helmi stream.
\end{abstract}




\section{Introduction} \label{sec:intro}

Stars with the lowest metal contents in their atmospheres are likely to be the direct descendants of the earliest stellar generations to be formed in the Universe \citep{Umeda+Nomoto03, bromm2004}. These stars can be seen as fossil records of the initial physical conditions of the Universe and provide important constraints to the formation and evolution of the Milky Way \citep[e.g.][]{Schorck+09, Li+10, Frebel+Norris15, Helmi+2020}, as well as its satellite galaxies \citep[e.g.][]{Kirby+08, Norris+08, Tolstoy2009, Frebel+10, simon2019}.

\begin{figure*}
 \includegraphics[width=\linewidth]{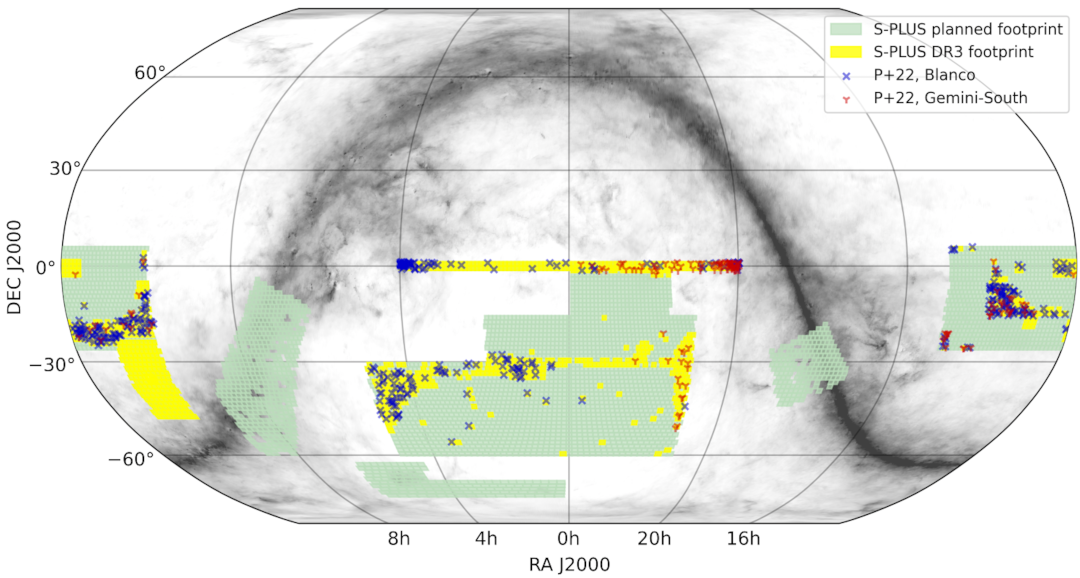}
 \caption{Sky distribution of the 522 stars in the P+22 sample, observed by Blanco (blue) and Gemini-South (orange). The S-PLUS planned footprint is shown in green and the S-PLUS DR3 is shown in yellow. The grey colourmap in the background corresponds to the extinction map of \citet{Schlegel+98}.}
\label{fig:radec}
\end{figure*}

The first efforts on building a large sample of metal-poor ([Fe/H]\footnote{\abund{A}{B} = $\log(N_A/{}N_B)_{\star} - \log(N_A/{}N_B) _{\odot}$, where $N$ is the number density of atoms $A$ and $B$ of a given element in the star ($\star$) and the Sun ($\odot$).}$\leq -1$) stars were based on the objective prism search technique, which was employed by the HK survey \citep{Beers+85, Beers+92} to identify around a thousand metal-poor candidates. A few years later, the Hamburg/ESO survey \citep{Christlieb+2008} was able to increase this number by an order of magnitude (see \citealp{Limberg+21} for a more recent calibration of these surveys). Large spectroscopic surveys, such as the Sloan Extension for Galactic Understanding and Exploration \citep[SEGUE][]{Yanny+2009}, the Large Sky Area Multi-Object Fiber Spectroscopic Telescope \citep[LAMOST][]{Cui+12, Luo+2015} and the Radial Velocity Experiment \citep[RAVE][]{Steinmetz+06, Kunder+17}, have also recently provided important results on the topic of metal-poor stars \citep[e.g.][]{Aoki+13, Lee+13, Roederer+2014, Placco+15, Placco+18, Li+2018, Yuan+2020, Aoki+2022}.

The advances in the determinations of atmospheric parameters directly from photometry \citep[e.g.][]{AllendePrieto2016, Casagrande+2010, Casagrande+2019, chiti2021} enables the study of different populations in our Galaxy using datasets containing millions of stars \citep[e.g.][]{Ivezic+2012, An+2013, An+2015}. Large photometric surveys, such as the Sloan Digital Sky Survey \citep[SDSS][]{York+2000, Eisenstein+15, Blanton+17}, SkyMapper \citep{Wolf+2018}, and Pristine \citep{Starkenburg+17}, are currently among the main sources for the discovery of new metal-poor candidates \citep{Huang+2022}. In particular, the strength of SkyMapper and Pristine in the search for metal-poor stars is the presence of a narrow-band filter around the \ion{Ca}{II} H and K absorption features, which are particularly sensitive to metallicity \citep{Bond1970, Bond1980, Beers+85}.

The new-generation multi-filter surveys J-PAS \citep[Javalambre Physics of the Accelerating Universe Astrophysical Survey;][]{Benitez+14,bonoli2021}, J-PLUS \citep[Javalambre Photometric Local Universe Survey;][]{Cenarro+2019}, and S-PLUS \citep[Southern Photometric Local Universe Survey;][]{MendesDeOliveira+19} also include a similar narrow-band filter around the \ion{Ca}{II} H and K lines, providing the necessary data for the search and characterization of stars in the  [Fe/H] $< -3$ regime. J-PAS plans to observe $\sim$8500  square degrees in the Northern hemisphere with a 2.5m telescope at the Observatorio Astrofísico de Javalambre, Spain, using 56 narrow-band filters. J-PLUS (which is an auxiliary survey to J-PAS, and uses an 80 cm telescope located on the same site) is covering an area of $\sim$8500 square degrees. using 12 filters: 7 narrow-band and 5 broad-band filters. The S-PLUS survey is the source of the data used in this work. It has instruments similar to those in J-PLUS, located in Cerro Tololo, Chile, and is observing $\sim$9000 square degrees in the Southern hemisphere.

The presence of the narrow-band filters is key in the determination of accurate and precise stellar parameters for both J-PLUS \citep{whitten2019, Galarza+22, Wang+22, Yang+22} and S-PLUS \citep{Whitten+21}. In particular, \citet[][hereafter, P+22]{Placco+22} analysed S-PLUS colour-colour diagrams and identified a region dominated by stars with \metal$\leq -1$: $(J0395-J0410)-(J0660-J0861) \in [-0.30:0.15]$ and $(J0395-J0660)-2\times(g-i) \in [-0.60:-0.15]$. P+22 selected 522 stars for medium-resolution spectroscopic follow-up and confirmed that 92\% of those stars were, indeed, metal-poor, including the ultra metal-poor star with the lowest measured carbon abundance at the time of its publication \citep{Placco+21}.

In this work, we characterize the P+22 sample in terms of isochronal stellar ages and kinematical properties. Our results allow us to better understand the stellar populations present in the sample and the nature of the non-metal-poor contaminants. Our main goal is to propose further restrictions to optimize the selection of the metal-poor candidates and, perhaps, even eliminate the need for medium-resolution spectroscopic follow-up to interpret the data in future studies with purely photometric stellar parameters. 

This paper is structured as follows. In Section~\ref{sec:sample} we describe the sample used in this work, while in Section~\ref{sec:Methodology} we formalize the isochronal method employed in the determination of stellar ages and describe the process of obtaining the kinematical parameters. Our results are presented in Section~\ref{sec:results} and discussed in Section~\ref{sec:discussion}, where several options are proposed to optimize the selection of metal-poor stars based on the correlations between the estimated parameters and the spectroscopic metallicity. Finally, our conclusions are presented in Section~\ref{sec:conclusions}.

\section{Sample} \label{sec:sample}

\subsection{The S-PLUS Data Release 3}

The photometry used in this paper comes from the third S-PLUS data release (DR3) of S-PLUS (in preparation), which covers an area of ${\sim}2000$ square degrees of the ${\sim}9000$ square degrees planned for the survey. S-PLUS photometric depths range from 19.1 to 20.5 mag (for a signal-to-noise threshold of 5) depending on the filter \citep{Almeida-Fernandes+22}. The sample used in this work\footnote{The data was obtained from the S-PLUS database (https://splus.cloud).} is restricted to magnitudes gSDSS $< 17.5$. In this range, the photometric errors are dominated by the uncertainty in the photometric calibration zero-points, estimated by \citet{Almeida-Fernandes+22} to be $\leq$ 10 mmags for filters J0410, J0430, gSDSS, J0515, rSDSS, J0660, iSDSS, J0861 and zSDSS; $\leq$ 15 mmags for filter J0378; and $\leq$ 25 mmags for filters uJAVA and J0395.

\begin{figure}
 \includegraphics[width=\columnwidth]{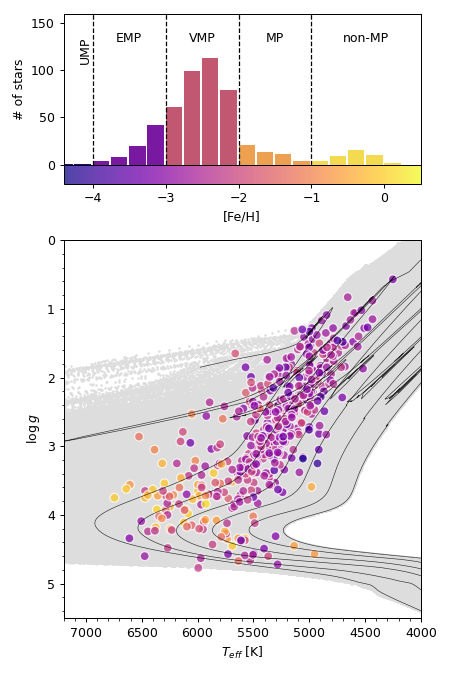}
 \caption{Kiel diagram (bottom) and metallicity distribution (top) of the 522 stars in the P+22 sample. The grey dots correspond to the MIST isochrones grid covering ages from 0.1 to 15 Gyr and metallicities from -4 to 0.5 dex. For reference, the solid black lines represent the 15 Gyr isochrones for metallicities -4, -2, -1, -0.5, 0, 0.5, from left to right.}
 \label{fig:sample_HR}
\end{figure}

\subsection{Low Metallicity Stars in S-PLUS DR3}

\citet[][]{Placco+22} identified a region dominated by metal-poor stars in S-PLUS colour-colour diagrams. After applying a few restrictions to the S-PLUS DR3 to ensure the quality of the photometry, P+22 selected 522 stars for medium-resolution spectroscopic follow-up: 384 stars were observed using COSMOS (Cerro Tololo Ohio State Multi-Object Spectrograph; \citealp{Martini+14}), installed in the Victor M. Blanco 4-meter telescope; and 138 stars were observed using GMOS (Gemini Multi-Objetc Spectrograph; \citealp{Davies+97, Gimeno+16}), in the Gemini-South telescope. Based on the aforementioned spectra, P+22 used the n-SSPP pipeline \citep{Beers+14, Beers+17} to measure stellar parameters for these stars.

In Figure \ref{fig:radec}, we show the sky distribution of the 522 stars in our sample, colour-coded according to the instrument used to measure the medium-resolution spectra (COSMOS in Blue and GMOS in red). The S-PLUS planned footprint is shown in green, while the area covered in DR3 is shown in yellow. As can be seen from the grayscale projection of the extinction map \citep{Schlegel+1998}, the stars in our sample are located at higher Galactic latitudes, $|b| > 15 \deg$, therefore, we expect our sample to be dominated by halo stars. 

As discussed in P+22, most of the selected stars are confirmed to be low-metallicity stars: 10\% are metal-poor (MP; $-2 <$ \metal $\leq -1$), 68\% are very metal-poor (VMP; $-3 <$ \metal $\leq -2$)  and 14\% are extremely metal-poor (EMP; $-4 <$ \metal $\leq -3$), with two stars being ultra metal-poor (UMP; $-5 <$ \metal $\leq -4$). The metallicity distribution in this sample is shown in the top panel of Figure~\ref{fig:sample_HR}. In the bottom panel of Figure~\ref{fig:sample_HR} we show the Kiel diagram of this sample, on top of the 
MIST (MESA Isochrones \& Stellar Tracks) isochrone set \citep{Dotter+2016}, represented in grey. For reference, we highlight (black lines) the isochrones of 15 Gyr for the metallicities  $-4$, $-2$, $-1$, $-0.5$, $0$ and $0.5$ (the lower metallicity isochrones have a higher temperature turn-off point). This diagram reveals that the sample is dominated by VMP stars in the giant branch stage, with most non-MP stars occupying the turn-off region. We note that there is good agreement between the sample distribution in this diagram and the region predicted by the MIST models for this metallicity range.

Even though P+22 estimate individual uncertainties for the parameters, we adopted more conservative uncertainties of 100~K, 0.35~dex and 0.20~dex for \teff, \logg, and \metal, respectively. These values are based on the estimations of \citet{Beers+14} for the n-SSPP pipeline used by P+22, divided by $\sqrt{2}$, to account for the fact that the estimations of \citet{Beers+14} arise from the comparison of the measurements between two different surveys.

\subsection{Gaia DR3}

A crossmatch between the P+22 sample and Gaia DR3 \citep{GaiaCollaboration+2022} is used in Section \ref{subsec:kinematics_orbital_integration} to integrate the Galactic orbits and estimate kinematical parameters. Gaia provides the necessary proper motions, line-of-sight velocities, and distances (in this case, we adopt the photo-astrometric distances from \citealp{Bailer-Jones+2021}). In total, 241 out of 522 stars in the P+22 sample have the 6D astrometric information necessary for the analysis (the bottleneck being the availability of line-of-sight velocities in the Gaia catalogue).

\section{Bayesian isochronal ages} 
\label{sec:Methodology}

The Bayesian approach to estimating isochronal ages has already been extensively discussed in the literature \citep{Pont+Eyer04, Jorgensen+Lindegren05, Takeda+07}. It is the best approach when only the atmospheric parameters are known (e.g. when no stellar rotation, chromospheric activity or asteroseismology data are available; \citealp{Soderblom10, Soderblom+14}), making it the best option for our sample.

In this work, we present a slight variation of the approach discussed in \citet[][hereafter, J\&L05]{Jorgensen+Lindegren05}. 
In J\&L05, the likelihood is only marginalized with respect to age, resulting in an age probability density function (pdf). In our method, we generalize this marginalization for any quantity that can be predicted by the stellar models (e.g. initial mass, age, metallicity, bolometric luminosity). 

J\&L05 consider that each point in the isochrone set can be represented by an age ($\tau$), metallicity ($\zeta$), and mass ($m$), and define the posterior probability function, $f(\tau, \zeta, m)$, in terms of the likelihood, $\mathcal{L}(\tau, \zeta, m)$ calculated from a set of observables ($\mathbf{q}$ and their respective uncertainties, $\sigma$), and the prior probabilities of each point in the isochrone $f_0(\tau, \zeta, m)$:

\begin{equation}
    f(\tau, \zeta, m) \propto \mathcal{L}(\tau, \zeta, m) f_0(\tau, \zeta, m)
\end{equation}

We propose a generalization of this equation for an arbitrary set of parameters that define a point in the isochrone grid (including, for example, the \afe\ abundance, and the $v/v_{\mathrm{crit}}$ stellar rotation\footnote{In this case $v_{\mathrm{crit}}$ corresponds to the velocity associated with the critical angular velocity that results in a net 0 gravity in the stellar surface \citep[see, for instance,][]{Gagnier+2019}.}), $\mathbf{h} = \{\tau, \zeta, m, [\alpha/\mathrm{Fe}], v/v_{\mathrm{crit}}, ...\}$. In this notation, for a given set of observables $\mathbf{q}$ (e.g. $\{T_{\mathrm{eff}}$, $\log g$, [Fe/H]$\}$), the multivariate pdf is given by:

\begin{equation}
    f({\mathbf{h}} | {\mathbf{q}}) \propto \mathcal{L}({\mathbf{q}} | {\mathbf{h}}) f_0({\mathbf{h}})
\end{equation}

In other words, \textbf{h} is the set of quantities that parametrize the stellar models, while \textbf{q} is the set of parameters we can measure independently from the stellar models. Therefore, we aim to use the isochronal method to infer one or more parameters in the set \textbf{h}, based on the measured values in the set \textbf{q}.

Assuming Gaussian uncertainties ($\sigma_i$) for the observable parameters, the likelihood $\mathcal{L}$ is given by:

\begin{equation}
\mathcal{L}({\mathbf{q}} | {\mathbf{h}}) = \left[ \prod_{i = 1}^{n} \frac{1}{(2\pi)^\frac{1}{2} \sigma_i} \right] \times \exp{\left[ \frac{\chi^2({\mathbf{q}} | {\mathbf{h}})}{2} \right]} 
\end{equation}

and
\begin{equation}
\label{eq::chi2}
\chi^2({\mathbf{q}} | {\mathbf{h}}) =  \sum_{i = 1}^{n} \left[ \frac{q_{i,\mathrm{obs}} - q_i(\mathbf{h})}{\sigma_i} \right]^2 
\end{equation}
where $q_{i,\mathrm{obs}}$ corresponds to the observed value for the observable $q_{i}$ and $\sigma_{i}$ is the corresponding uncertainty for this value. The parameter $q_i(\mathbf{h})$ represents the observable value predicted by the model defined by the parameter set \textbf{h}.

We adopt uniform distributions for the priors in all isochrone parameters, except for the mass. The limits for these uniform distributions are set by the range of the models: constraining the metallicity between $-4.0$ and $0.5$ (the full range available for the MIST isochrones), and the ages between 0.1 and 15 Gyr (consistent with the values adopted by J\&L05). The prior can then be written as $f_0({\mathbf{h}}) = \xi(m)$, where $\xi$ corresponds to the initial mass function (IMF) and is implied not to be dependent on the other parameters. Here we adopt the \citet{2002SciKroupa2002} IMF.

Finally, we obtain the pdf for any parameter by numerically marginalizing the multivariate pdf. In this work, we simply replace the integral with a Riemann sum.  For example, for the age ($\tau$), the pdf can be obtained by:
\begin{equation} \label{eq::marginalization}
    f(\tau_j) \delta \tau \propto \sum_{\tau \in [\tau_j, \tau_j + \delta \tau[}{ f\left({\mathbf{h}} | \mathbf{q}\right) \times D(\mathbf{h})}  
\end{equation}
where $D(\mathbf{h})$ represents the n-dimensional volume in the parameter space $\mathbf{h}$ and is calculated according to the difference between each consecutive value in the model grid:
\begin{equation} \label{eq1}
\begin{split}
   D(\mathbf{h}_i = \{\tau, \zeta, m, ...\}_i) = & (\tau_{i+1} - \tau_{i}) \times \\ & (\zeta_{i+1} - \zeta_{i}) \times \\ & (m_{i+1} - m_{i}) \times ...
\end{split}
\end{equation}

The main advantage of this approach over the method proposed by J\&L05 is that any quantity predicted by the isochrones can be used in place of $\tau$ in Equation~\ref{eq::marginalization}. Therefore, this same set of equations can be used to obtain pdfs for atmospheric parameters (e.g. [Fe/H], $T_{\mathrm{eff}}$, $\log g$), the initial mass ($m$), and even the absolute magnitude in a given passband.

\subsection{Point Estimation}

The Bayesian approach results in a pdf. Even though a point estimation from the pdf implies loss of information, it is an adequate approach to explore correlations between the properties of the sources in a given dataset. J\&L05 discuss the use of the mean, the median or the mode for the age characterization. We choose not to use the mode for the analysis because it is limited by the intervals in the grid of the models, which in our case are steps of 0.2 Gyr. The mean and the median provide very similar results (the average differences being of the order of 1\% of the ages). For the subsequent analysis, we adopt the median as the point estimator for the age and the percentiles of 16\% and 84\% as the lower and upper limits (which corresponds to one standard deviation from the median for a normal distribution).

\subsection{Isochrone set}

The isochrone set used in this work to realize the function $q_i(\mathbf{h})$, in Equation~\ref{eq::chi2}, was obtained from the MIST database \citep{Dotter+2016}. We have used version 1.2 and chosen the rotation $v/v_{\mathrm{crit}}=0.4$ because the models with no stellar rotation are not complete for the lower metallicity range (\metal $< -2$). All the models have solar-scaled abundances, and our grid ranges from 0.1 to 15.0 Gyr in steps of 0.2 Gyr in ages and from $-$4.0 to $+$0.5~dex in steps of 0.1~dex in metallicities. The full grid of models is shown in Figure~\ref{fig:sample_HR} as grey circles.

\begin{figure}
 \includegraphics[width=\columnwidth]{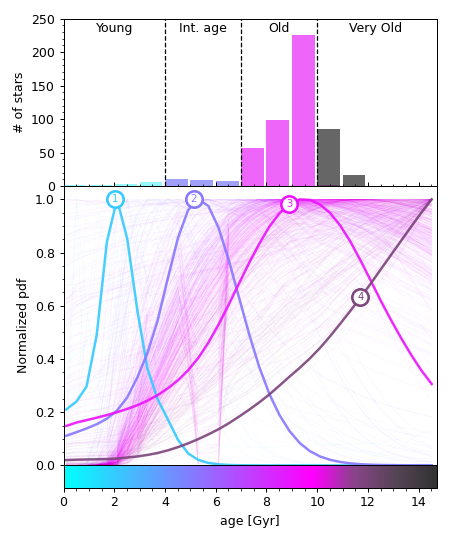}
 \caption{Age pdfs obtained for the 522 P+22 sample using the isochronal method (bottom), with four representative stars highlighted (see Table~\ref{tab:ages}). The top panel shows the distribution of the characterized median ages, coloured according to the classified age group (see Table \ref{tab:groups}).}
 \label{fig:ages_hist}
\end{figure}

\subsection{Age-mass degeneracy}

We also divide the age determination into two steps: 1) we characterize the stellar initial mass ($m_{\mathrm{init}}$) from the mass pdf obtained by using the known stellar parameters as the input; 2) secondly, we repeat the analysis including the predicted mass among the inputs, and only then we characterize an age from the resulting age pdf. The age pdf obtained in step 2 is significantly smoother, favouring the characterization of a single age from the median of the distribution. Even though we do not use for the analysis the age pdf obtained in step 1, we note that the difference in ages between steps 1 and 2 are minimal: $\approx$2.5\% for stars with $m_{\mathrm{init}} < 0.8 \, M_\odot$ (17\% of the sample), and $\approx$4.3\% for stars with $m_{\mathrm{init}} \geq 0.8 \, M_\odot$ (88\% of the sample).

A comparison between ages determined by our method and those derived using the J\&L05 method for the Geneva-Copenhagen Survey \citep{Nordstrom+04, Casagrande+11} is presented in Appendix~\ref{sec:verification_isochronal}.

\section{Results} 
\label{sec:results}

\subsection{Ages}
\label{subsec:ages}

\begin{table*}
\caption{Ages and initial mass for the four stars shown in Figure \ref{fig:ages_hist} and the first six entries of the catalogue. Values were estimated using the atmospheric parameters ($T_{\mathrm{eff}}$, {[}Fe/H{]}, $\log g$) from P+22 as inputs for the isochronal method. The value of $\tau_\mathrm{ML}$ is the most likely age extracted from the resulting age pdf, $\tau_\mathrm{E}$ is the expected age, while the numbers indicate the corresponding percentiles of this distribution. The adopted age corresponds to the median age, with uncertainties estimated according to the difference between this age and the 16\% and 84\% percentiles. The full table is only available in electronic format.}
\label{tab:ages}
\begin{tabular}{lcccccccccc}
\hline \hline                                                                   & $T_{\mathrm{eff}}$ & {[}Fe/H{]} & $\log g$ & $m_{\mathrm{init}}$ & $\tau_\mathrm{ML}$ & $\tau_{E}$ & $\tau_{50}$ & $\tau_{16}$ & $\tau_{84}$ & $\tau_{\mathrm{adop}}$ \\
ID & [K] & dex & dex & [$M_\odot$] & [Gyr] & [Gyr] & [Gyr] & [Gyr] & [Gyr] & [Gyr] \\
\hline
SPLUS J200116.81-011625.9 & $6638$ & $-0.10$ & $3.62$ & $1.253$ & $2.110$ & $2.301$ & $2.016$ & $1.159$ & $3.044$ & $2.016^{+1.028}_{-0.857}$ \\[+0.07in]
SPLUS J035659.53+000841.9 & $6606$ & $-0.73$ & $3.56$ & $0.952$ & $5.310$ & $5.266$ & $5.127$ & $3.237$ & $6.876$ & $5.127^{+1.749}_{-1.890}$ \\[+0.07in]
SPLUS J035546.72+002806.4 & $5985$ & $-0.42$ & $3.72$ & $0.863$ & $9.310$ & $8.765$ & $8.897$ & $5.256$ & $11.850$ & $8.897^{+2.953}_{-3.641}$ \\[+0.07in]
SPLUS J035508.83+001433.9 & $5964$ & $-2.53$ & $3.42$ & $0.809$ & $14.510$ & $11.200$ & $11.700$ & $8.137$ & $13.730$ & $11.700^{+2.030}_{-3.563}$ \\[+0.07in]
SPLUS J000445.50+010117.0 & $5227$ & $-2.37$ & $2.56$ & $0.868$ & $8.910$ & $9.441$ & $9.356$ & $5.756$ & $12.780$ & $9.356^{+3.424}_{-3.600}$ \\[+0.07in]
SPLUS J001736.44+000921.7 & $4993$ & $-2.63$ & $2.19$ & $0.878$ & $10.110$ & $9.472$ & $9.417$ & $5.803$ & $12.770$ & $9.417^{+3.353}_{-3.614}$ \\[+0.07in]
SPLUS J002554.41-305032.0 & $5186$ & $-2.21$ & $1.72$ & $0.915$ & $5.710$ & $7.999$ & $7.533$ & $4.127$ & $11.750$ & $7.533^{+4.217}_{-3.406}$ \\[+0.07in]
SPLUS J002712.10-313352.1 & $5257$ & $-2.27$ & $2.74$ & $0.857$ & $8.910$ & $9.510$ & $9.433$ & $5.826$ & $12.840$ & $9.433^{+3.407}_{-3.607}$ \\[+0.07in]
SPLUS J002712.43+010037.0 & $5394$ & $-2.29$ & $3.41$ & $0.813$ & $14.510$ & $10.410$ & $10.690$ & $6.916$ & $13.410$ & $10.690^{+2.720}_{-3.774}$ \\[+0.07in]
SPLUS J003555.86-420431.0 & $5645$ & $-2.53$ & $3.38$ & $0.824$ & $14.510$ & $10.700$ & $11.130$ & $7.181$ & $13.590$ & $11.130^{+2.460}_{-3.949}$ \\[+0.07in] \hdashline
\end{tabular}
\end{table*}

\begin{figure}
 \includegraphics[width=.99\columnwidth]{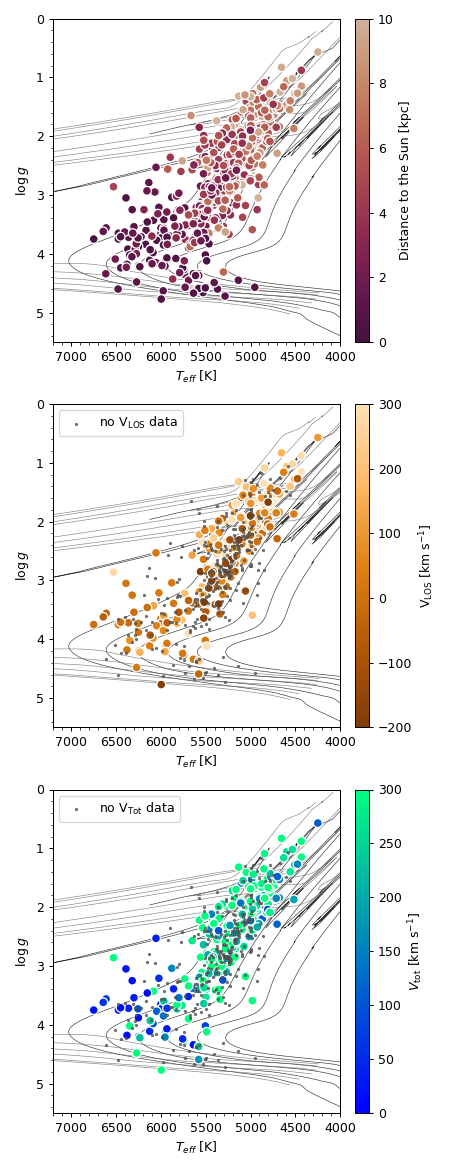}
 \caption{Kiel diagram for the same stars as those in Figure~\ref{fig:sample_HR}, color-coded by Gaia's EDR3 distances (top), Gaia's DR3 $V_{\mathrm{LOS}}$ (middle) and total velocities (bottom). The solid lines represent the 0.1 Gyr (grey) and 15 Gyr (black) isochrones for metallicities $-$4, $-$2, $-$1, $-$0.5, 0.0, $+$0.5.}
 \label{fig:hess_diagram_gaia}
\end{figure}

We applied the isochronal method described in Section \ref{sec:Methodology} to characterize the ages for all stars in the P+22 sample. The input parameters were the effective temperature, metallicity, and surface gravity ($\mathbf{h} = \{T_{\mathrm{eff}}, \mathrm{[Fe/H]}, \log g\}$). The pdfs for all 522 stars in the sample are shown in the bottom panel in Figure~\ref{fig:ages_hist}.

The stars are distributed across the entire range of ages, with the most likely values varying between 0.11 and 14.51 Gyr. The pdfs vary significantly with age, with the interquartile difference being much smaller for younger stars. This can be seen clearly in the four highlighted pdfs in Figure \ref{fig:ages_hist}. The increase in interquartile range is the effect of the change in isochrone spacing of consecutive isochrones between the turn-off and the giant branch regions. 

In Table \ref{tab:ages} we show the characterized ages and percentiles for the four highlighted stars in Figure~\ref{fig:ages_hist}. The full distributions of median ages are shown in the top panel of Figure~\ref{fig:ages_hist}. The majority (94 \%) of the stars are older than 6 Gyr. The upper and lower limits are obtained from the difference between the median and the 16\% and 84\% percentiles.

After characterizing the ages, we separate the stars in four different groups: Young ($\tau_{\mathrm{adop}} \leq 4$ Gyr), Intermediate age ($4 \mathrm{Gyr} < \tau_{\mathrm{adop}} \leq 7$ Gyr), Old ($7 \mathrm{Gyr} < \tau_{\mathrm{adop}} \leq 10$ Gyr), and Very Old ($\tau_{\mathrm{adop}} > 10$ Gyr)

\subsection{Kinematical parameters and orbital integration} \label{subsec:kinematics_orbital_integration}

In order to study the kinematical properties of the stars in our sample, we performed a cross-match (with a tolerance of 3 arcsec) between the stars in P+22 and Gaia's DR3 \citep{GaiaCollaboration+2022}. We use the photo-astrometric distances provided by \citet{Bailer-Jones+2021}, while adopting the coordinates, proper motions, and radial velocities from  Gaia DR3.  We have chosen to use only the Gaia DR3 radial velocities as they are more reliable than the ones obtained by P+22 using the n-SSPP pipeline. In particular, the dispersion between Gaia's and other similar-resolution spectroscopic $V_\mathrm{LOS}$ has been found to be $\approx18$ km s$^{-1}$, as noted by \citet{Limberg+21} and \citet{Shank+2022}.

The orbital parameters were estimated by integrating the orbits for 10 Gyr (forward) in the potential described in \citet{McMillan2017}, using the \texttt{Galpy} python package \citep[][]{Bovy+15}. We adopted $R_\odot = 8.21$ kpc, $V_\odot = 233.1$ km s${-1}$ \citep{McMillan2017}, and $(U_\odot, V_\odot, W_\odot) = (11.1, 12.24, 7.25)$ km s$^{-1}$ from \citet{Schonrich+2010}. We also used \texttt{Galpy} to calculate the cartesian $U$, $V$, $W$ velocities\footnote{By definition in \texttt{Galpy}, $U$ is positive towards the Galactic center.}, and the corresponding total velocity: 
\begin{equation}
 V_\mathrm{Tot} = \left(U^2 + V^2 + W^2 \right)^{\frac{1}{2}}   
\end{equation}

In Figure~\ref{fig:hess_diagram_gaia} we show the Kiel diagram for the P+22 sample, colour-coded by heliocentric distance (top), line of sight velocity ($V_\mathrm{LOS}$; middle), and total velocity ($V_\mathrm{Tot}$; bottom). We observe a clear correlation between distances and position in the Kiel diagram, with the closest stars located in the turn-off region and the more distant stars located in the giant branch. This is the result of the restricted magnitude range covered by the P+22 sample, where most of the stars have apparent magnitudes ($r\mathrm{SDSS}$) between 13 and 16. The line-of-sight velocity distribution is evenly scattered: there are missing values for both the turn-off and the giant branch stars, so we do not expect a bias in terms of stellar class to be introduced by the availability of these velocities. These missing values also translate to the total velocity, which could only be calculated for 241 out of 522 stars (46\%). In this case, we see that the turn-off is dominated by low-velocity stars, while the giant branch concentrates most of the high-velocity stars in the sample.

\begin{figure}
 \includegraphics[width=\columnwidth]{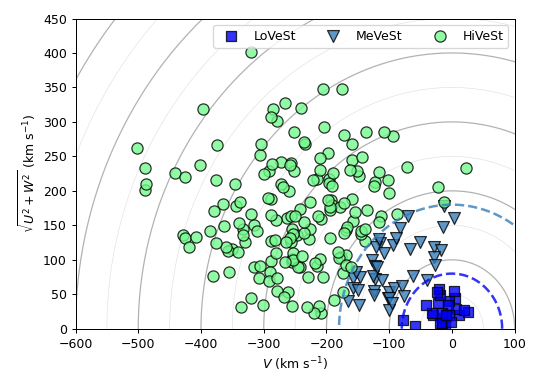}
 \caption{Toomre diagram for the 241 P+22 stars with 6D astrometric parameters available in Gaia's DR3. Stars were classified according to their total velocity: Low velocity stars with $V_\mathrm{Tot} \leq 80 \mathrm{km} \, \mathrm{s}^{-1}$ (blue squares); Medium velocity stars, with $80 \mathrm{km} \, \mathrm{s}^{-1} < V_\mathrm{Tot} \leq 180 \mathrm{km} \, \mathrm{s}^{-1}$ (dark cyan triangles); and High velocity stars $V_\mathrm{Tot} > 180 \mathrm{km} \, \mathrm{s}^{-1}$ (green circles).}
 \label{fig:uvw}
\end{figure}

We characterized the apogalactic and perigalactic radius ($R_\mathrm{apo}$ and $R_\mathrm{peri}$, respectively) and the maximum distance above the Galactic plane ($z_{\mathrm{max}}$). We use these parameters to calculate the eccentricity ($\ecc$), and the 'diskness', a quantity proposed by \citep{Sales-Silva2019}:

\begin{equation}
    \ecc = \frac{R_{\mathrm{apo}} - R_{\mathrm{peri}}}{R_{\mathrm{apo}} + R_{\mathrm{peri}}}
\end{equation}

\begin{equation}
    \mathrm{diskness} = \frac{R_{\mathrm{apo}} - z_{\mathrm{max}}}{R_{\mathrm{apo}} + z_{\mathrm{max}}}
\end{equation}

Additionally, the orbital energy and angular momentum are also computed and included in our final catalogue.

\begin{table*}
\caption {UVW velocities and orbital parameters estimated for the four stars shown in Figure \ref{fig:ages_hist} and the first six entries of the catalogue. The data for the complete sample, including additional dynamical parameters, can be found in electronic format.}
\resizebox{\linewidth}{!}{%
\begin{tabular}{lcccccccccc} \hline \hline
ID                        & \begin{tabular}[c]{@{}c@{}}U\\ {[}\kms{]}\end{tabular} & \begin{tabular}[c]{@{}c@{}}V\\ {[}\kms{]}\end{tabular} & \begin{tabular}[c]{@{}c@{}}W\\ {[}\kms{]}\end{tabular} & \begin{tabular}[c]{@{}c@{}}\Rapo\\ {[}kpc{]}\end{tabular} & \begin{tabular}[c]{@{}c@{}}\Rperi\\ {[}kpc{]}\end{tabular} & \begin{tabular}[c]{@{}c@{}}\zmax\\ {[}kpc{]}\end{tabular} & \ecc  & diskness & \begin{tabular}[c]{@{}c@{}}$E / 10^5$\\ {[}km$^2$ s$^{-2}${]}\end{tabular} & \begin{tabular}[c]{@{}c@{}}$L_Z / 10^3$\\ {[}kpc km s$^{-1}${]}\end{tabular} \\ \hline
SPLUS J200116.81-011625.9 & $-53.9$ & $-23.9$ & $-0.1$ & $8.51$ & $5.49$ & $0.46$ & $0.22$ & $0.90$ & $-1.65$ & $1.55$ \\ 
SPLUS J035659.53+000841.9 & $70.4$ & $-39.0$ & $3.5$ & $10.76$ & $6.09$ & $0.60$ & $0.28$ & $0.89$ & $-1.54$ & $1.82$ \\ 
SPLUS J035546.72+002806.4 & $-14.8$ & $-41.5$ & $31.0$ & $8.79$ & $7.01$ & $0.96$ & $0.11$ & $0.80$ & $-1.58$ & $1.79$ \\ 
SPLUS J035508.83+001433.9 & --- & --- & --- & --- & --- & --- & --- & --- & --- & --- \\ 
SPLUS J000445.50+010117.0 & $62.1$ & $-157.0$ & $39.9$ & $9.39$ & $2.44$ & $2.85$ & $0.59$ & $0.53$ & $-1.70$ & $0.85$ \\ 
SPLUS J001736.44+000921.7 & $-122.8$ & $-215.3$ & $178.9$ & $11.03$ & $9.22$ & $11.03$ & $0.09$ & $0.00$ & $-1.45$ & $-0.00$ \\ 
SPLUS J002554.41-305032.0 & $132.3$ & $-122.3$ & $166.2$ & $18.64$ & $3.04$ & $12.65$ & $0.72$ & $0.19$ & $-1.37$ & $0.92$ \\ 
SPLUS J002712.10-313352.1 & $102.5$ & $-292.0$ & $-205.3$ & $10.21$ & $6.10$ & $9.92$ & $0.25$ & $0.01$ & $-1.55$ & $-0.38$ \\ 
SPLUS J002712.43+010037.0 & $43.6$ & $-381.3$ & $-63.2$ & $9.14$ & $3.61$ & $2.40$ & $0.43$ & $0.58$ & $-1.68$ & $-1.11$ \\ 
SPLUS J003555.86-420431.0 & $117.6$ & $-51.1$ & $45.2$ & $12.03$ & $4.42$ & $2.41$ & $0.46$ & $0.67$ & $-1.54$ & $1.49$ \\ \hdashline
\end{tabular}%
}
\end{table*}

The Toomre diagram for the stars in this sample is shown in Figure~\ref{fig:uvw}. We classify the stars in three groups: (i) low velocity stars (LoVeSt) if their $V_\mathrm{Tot}$ is lower than 80 $\mathrm{km} \mathrm{s}^{-1}$; (ii) medium velocity star (MeVeSt) if $V_\mathrm{Tot}$ is between 80 and 180 $\mathrm{km} \mathrm{s}^{-1}$; and (iii) high velocity star (HiVeSt) if $V_\mathrm{Tot}$ is higher than 180 $\mathrm{km} \mathrm{s}^{-1}$. The sample is dominated by high-velocity stars (72\% of the stars with known $V_\mathrm{Tot}$).

We associate the LoVeSt with the thin disk population. In the case of MeVeSt, while they have kinematical properties compatible with the canonical thick disk, these stars, by construction, have much lower metallicity than expected for the thick disk population. Nonetheless, VMP stars have already been discovered in this population (see, for example, \citealp{Sestito+2020, DiMatteo+2020, Cordoni+2021, Limberg+21b}). These VMP thick-disk stars are likely associated with the so-called 'metal-weak' thick disk (see \citealp{Morrison+1990} for an early discussion). For the HiVeSt, a more detailed analysis in the integrals of motion space is necessary to differentiate halo stars formed in situ from those associated with a past accretion event (see Section \ref{sec:substructures}).

\begin{figure*}
 \includegraphics[width=\linewidth]{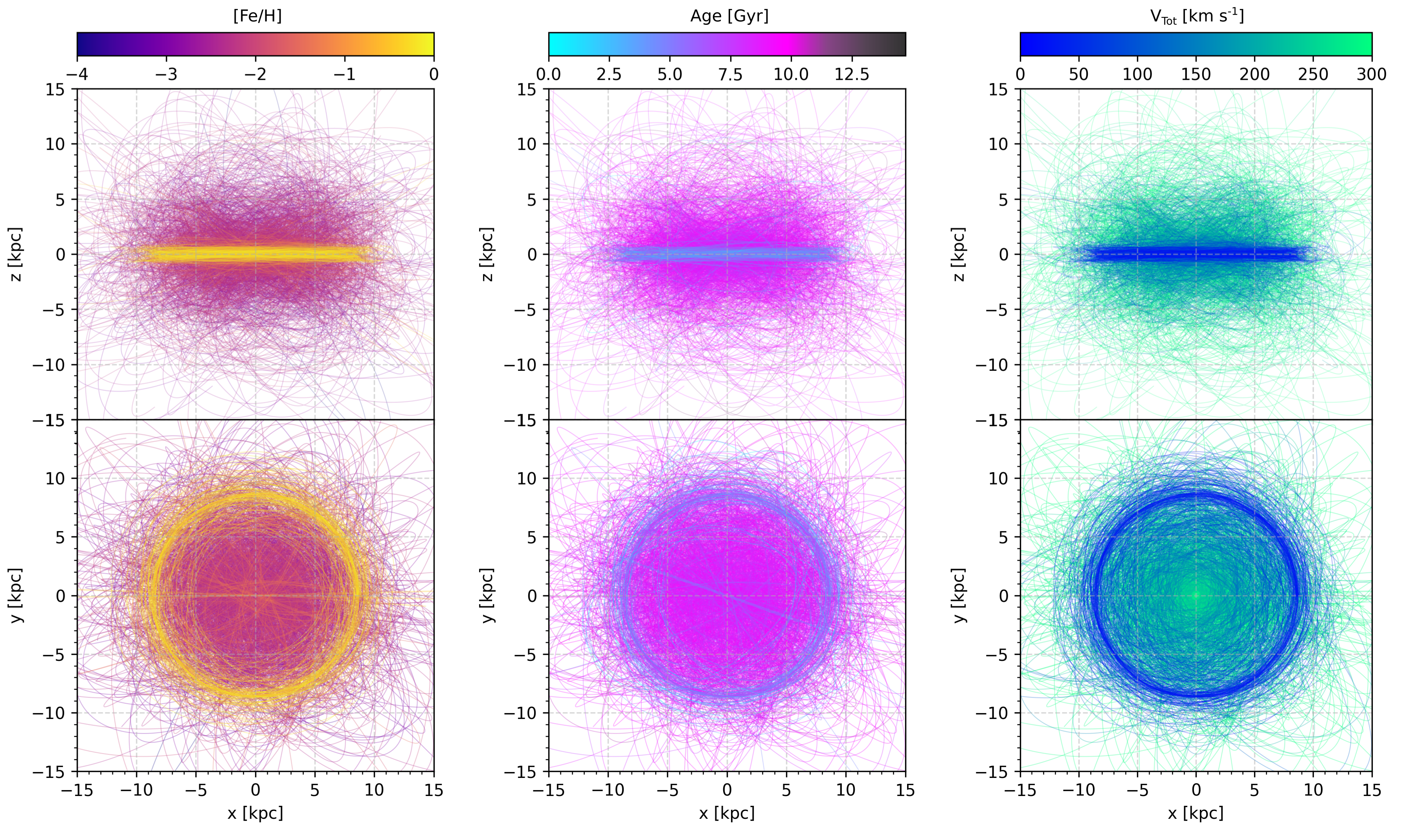}
 \caption{Resulting trajectories from the orbital integration for the 241 stars in the P+22 sample with known 6D astrometric parameters, shown in XYZ cartesian coordinates. Only the first billion years of the trajectories are shown.  The orbits are coloured according to the stellar metallicity (left), age (middle) and $V_{\mathrm{Tot}}$ (right).
 }
 \label{fig:example_orbits}
\end{figure*}

\begin{figure}
 \includegraphics[width=.9\columnwidth]{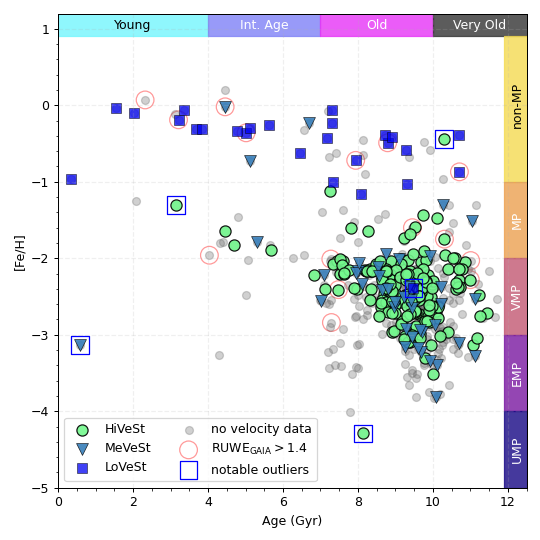}
 \caption{Correlation between age and metallicity for the stars in the P+22 sample. Stars are divided into velocity groups: LoVeSt (blue squares), MeVeSt (dark cyan triangles) and HiVeSt (green circles). Stars with no velocity data available are shown as grey circles. Age and metallicity groups are also indicated in the borders of the plot. Notable outliers are marked with a blue open square and discussed in Section \ref{sec:outliers}.}
 \label{fig:ages_feh_kin}
\end{figure}

In Figure~\ref{fig:example_orbits} we show the results of the orbital integration of the stars in the P+22 sample. The trajectories were integrated for 10 Gyr (for simplicity, only the first billion years are represented in the Figure) and are shown in Galactic Cartesian coordinates XYZ. The orbits are coloured according to metallicity (left), age (middle) and $V_\mathrm{Tot}$ (right). We can see most of the non-MP contaminants in the sample, as well as young and intermediate-age stars, correspond to the thin disk population. We also observe that most of the metal-poor stars have trajectories consistent with the thick disk and halo populations.

\section{Discussion} \label{sec:discussion}

Our main goal in characterizing the chemodynamical properties and ages of the P+22 sample is to understand the different stellar populations present in our dataset and use this information to increase the purity and accuracy in selecting metal-poor stars in S-PLUS. We accomplish this by dividing the sample into different subclasses of metallicity, age, and $V_\mathrm{Tot}$. For the metallicity, we follow the definitions of \citet{Beers+Christlieb05}. The velocities are divided into LoVeSt, MeVeSt and HiVeSt, as previously discussed in Section~\ref{subsec:kinematics_orbital_integration}. The ages are separated into Young, Intermediare age, Old and Very Old groups, as defined in Section \ref{subsec:ages}. The definition of these groups is summarized in Table~\ref{tab:groups}.

\begin{table}
\centering
\caption{Summary of definitions and acronyms of the different groups defined for metallicity, total velocity, and age.}
\label{tab:groups}
\resizebox{\columnwidth}{!}{%
\begin{tabular}{lclc} \hline \hline
Nomenclature          & Abbr. & Definition & Fraction                                                 \\ \hline
\multicolumn{4}{c}{\textbf{Metallicity}}                                                                    \\ \hline
Non-metal-poor        & non-MP       & {[}Fe/H{]} $> -1$ & 8\%                                          \\
Metal-poor            & MP           & $-2 <$ {[}Fe/H{]} $\leq -1$  & 10\%                               \\
Very metal-poor       & VMP          & $-3 <$ {[}Fe/H{]} $\leq -2$  & 68\%                               \\
Extremely metal.poor  & EMP          & $-4 <$ {[}Fe/H{]} $\leq -3$ & 15\%                                \\
Ultra metal-poor      & UMP          & {[}Fe/H{]} $\leq -4$ & 0\%                                       \\ \hline
\multicolumn{4}{c}{\textbf{Total Velocity}}                                                                     \\ \hline
Low velocity star     & LoVeSt          & $V_{\mathrm{tot}} \leq 80$ km s$^{-1}$  & 5\%                    \\ 
Medium velocity star  & MeVeSt          & $80$ km s$^{-1} < V_{\mathrm{tot}} \leq 180$ km s$^{-1}$ & 8\% \\
High velocity star    & HiVeSt          & $V_{\mathrm{tot}} > 180$ km s$^{-1}$ & 33\%                       \\ 
No velocity data & & & 54\% \\ \hline
\multicolumn{4}{c}{\textbf{Age}}                                                                            \\ \hline
Young            &              & Age $\leq 4$ Gyr  & 2\%                                          \\
Intermediate Age &  Int. Age      & $4$ $<$ Age (Gyr) $\leq 7$  & 5\%                              \\
Old              &              & $7$ $<$ Age (Gyr) $\leq 10$ & 73\%                              \\
Very Old         &              & Age $> 10$ Gyr & 20\%       \\\hline \hline                                     
\end{tabular}%
}
\end{table}

\subsection{Age-Metallicity relation} \label{subsec:AMR}

The age-metallicity relation for the 522 stars in our sample is shown in Figure~\ref{fig:ages_feh_kin}. A very clear distinction is seen between the LoVeSt and the other two kinematical groups. The ages are uniformly distributed in the LoVeSt group, and most of the stars in this group are non-MP (considered contaminants in this sample that targets metal-poor stars). Out of the three exceptions, only one is a significant outlier (SPLUS J110831.29-223514.5) and is discussed in Section~\ref{sec:outliers}, together with other notable outliers. We verified that only 15 stars have Gaia RUWE\footnote{Gaia's Renormalised Unit Weight Error (RUWE) is expected to be around 1.0 for single stars} greater than 1.4 (red circles in Figure~\ref{fig:ages_feh_kin}) and their distribution in the considered parameter spaces is apparently random, thus we do not expect our results to be biased due to non-resolved binary stars.

We note the presence of non-MP stars with velocities typical of the thin disc and with ages greater than 8 Gyr. These stars are good candidates for old thin disk stars that formed in the inner Galaxy and migrated to the Solar neighbourhood (thus would have higher [Fe/H] than stars forming at the same time in the Solar neighbourhood). A similar population is found by \citet{BeraldoESilva+2021} in a sample of old stars.

The MeVeSt and HiVeSt groups are dominated by old and very old VMP stars. The majority of the stars in the MeVeSt group are older than 9 Gyr, as expected for the thick disk \citep{Miglio+2021}. The lack of VMP, EMP and UMP stars in the Young and IntAge groups is coherent with our current understanding of the chemical evolution of the Galaxy \citep[e.g.][]{Starkenburg+2017b, ElBadry+2018}.

Both for the LoVeSt and HiVeSt there appears to be a metallicity-age gradient. However, we do not attempt to fit these gradients, as they would not be representative of the global properties of the Galaxy considering the strong selection effects imposed by the colour constraints used in the P$+$22 sample selection. Furthermore, this correlation could also be caused by the degeneracy between age and metallicity when applying the isochronal method.

This analysis by itself already shows that the main source of non-MP contaminants in the sample are the LoVeSt, attributed to the thin disk population, where metal-rich stars are much more predominant than in the halo or the thick disk \citep[e.g.][]{Haywood+13}. 

\subsection{Colour-colour diagrams}

\begin{figure*}
 \includegraphics[width=.95\linewidth]{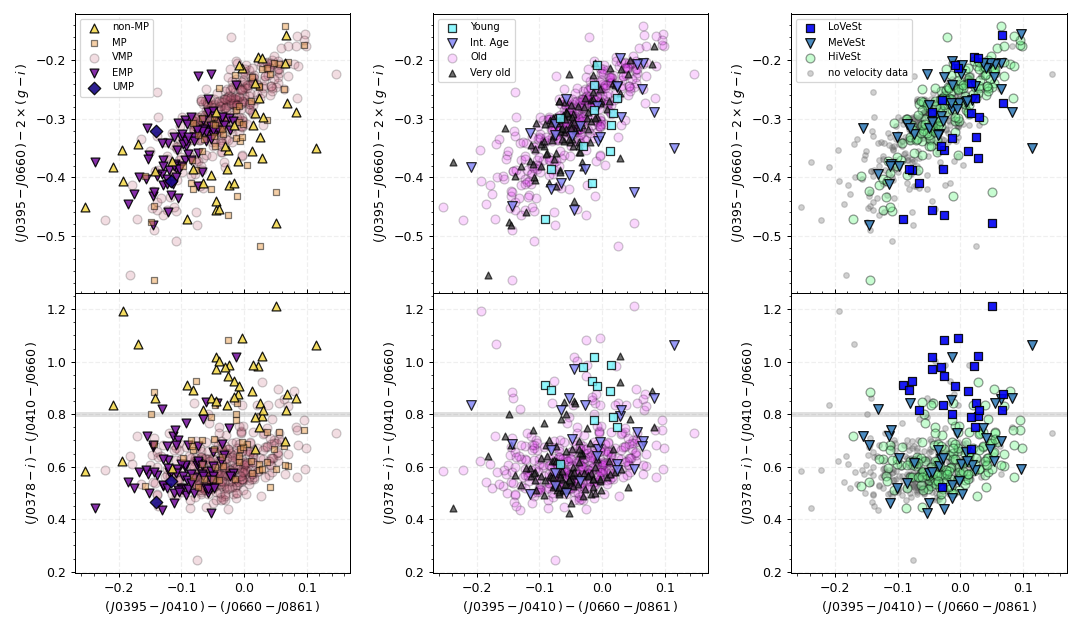}
 \caption{Pseudo-color diagrams calculated from S-PLUS DR3 photometry divided in subgroups of metallicity (left), age (middle) and total velocity (right). The top panels show the pseudo-colours diagram used in the original selection of the metal-poor stars by P+22, while the bottom panels display the pseudo-colour cut proposed to optimize the selection of metal-poor candidates.}
 \label{fig:color-color}
\end{figure*}

\begin{figure*}
 \includegraphics[width=.95\linewidth]{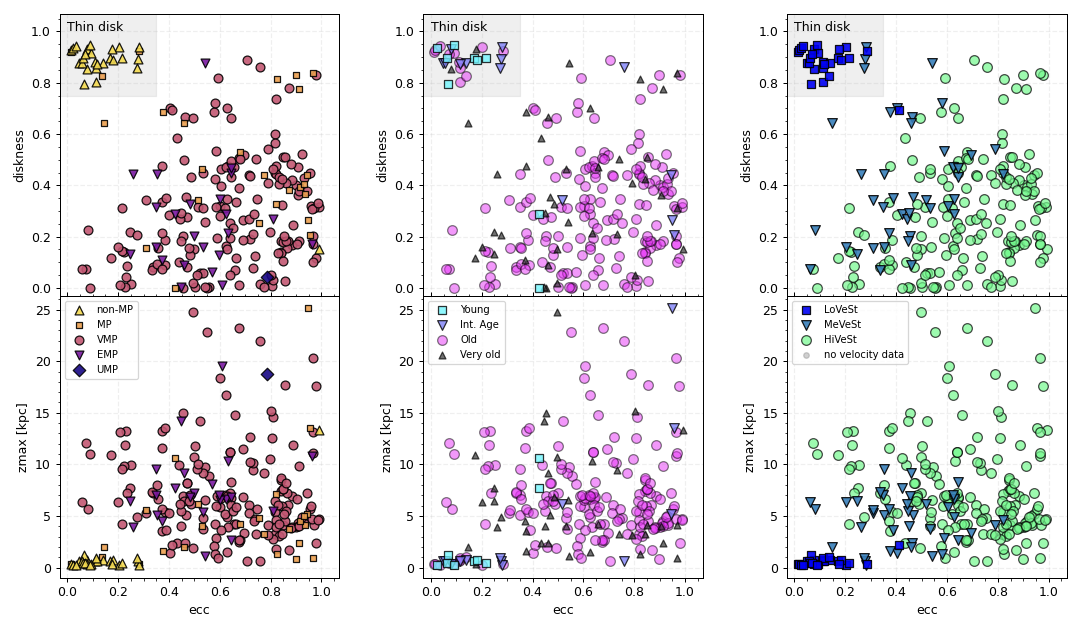}
 \caption{$\ecc$-diskness diagram (top) and $\ecc$-$z_\mathrm{max}$ diagram (bottom) for the stars in the P+22 sample, divided into groups of metallicity (left), age (middle) and total velocity (right). The shaded region in the top diagram is the locus of the thin disk stars, as identified by \citet{Sales-Silva2019} using data from the Geneva-Copenhagen survey.}
 \label{fig:orbital_params}
\end{figure*}

We now analyse the distribution of the metallicity, age, and velocity groups in the colour-colour diagram ($(J0395-J0410)-(J0660-J0861)$ x $(J0395-J0660)-2\times(g-i)$) used by \citet{Placco+22} to select the metal-poor candidates in the S-PLUS data (top panels in Figure \ref{fig:color-color}). Our goal is to investigate if the locus of each group indicates the possibility of further restricting the selection in order to increase the accuracy in identifying metal-poor candidates.

When dividing the sample by metallicity, we see a significant overlap between the groups, but a difference in terms of location and dispersion is noticeable in these parameter spaces. The MP and VMP stars are scattered over the whole diagram, but a clear density peak is observed around coordinates ($-0.02$, $-0.3$). The non-MP contaminants are also scattered throughout the diagram, and no clear peak in the density distribution is observed. This makes it unfeasible to remove these contaminants by further restricting the colour cuts. We see, however, that it is possible to significantly increase the ratio of EMP and UMP stars in the selection, as all of these stars are located within ([$-0.3$, 0], [$-0.5$, $-0.2$]). For instance, applying this cut to the P+22 sample brings the fraction of EMP to 18.9\% (in contrast with the original 14.4\% fraction).

We do not observe a significant correlation between age and locus in the colour-colour diagram shown in the middle-top panel of Figure \ref{fig:color-color}. In the case of the velocity groups (top right panel), the distributions of MeVeSt and HiVeSt are very similar, while the LoVeSt distribution is slightly offset to the right. This is because none of the EMP stars in our sample is LoVeSt, and this metallicity group dominates the distribution in the $[(J0395-J0410)-(J0660-J0861)] < -0.1$ region.

The selection effectiveness changes for the colour-colour diagram $(J0395-J0410)-(J0660-J0861)$ $\times$ $(J0378-i)-(J0410-J0660)$. This diagram was proposed by \citet{Placco+22} to further clean the sample from the non-MP contaminants. Here we see that selecting stars with $[(J0378-i)-(J0410-J0660)] < 0.8$ not only eliminates most of the non-MP contaminants, but also alters the age distribution by removing most of the young stars, and the velocity distribution by eliminating most of the LoVeSt. These results also suggest that a selection using spatial velocities (or orbital parameters) is able to produce similar results as these colour cuts. This approach has already been proposed in the literature \citep{Placco+18, Limberg+21b} and here it is explored in Section~\ref{sec:selection_hists}.

\subsection{Dynamical properties}

We used the results of the orbital integration in order to further analyse the properties of the different stellar groups present in the sample (for the stars with 6D astrometric data available). We focus our discussion on the $\ecc$-$z_\mathrm{max}$ diagram, as well as on the $\ecc$-diskness diagram, which was shown by \citet{Sales-Silva2019} to be a good diagnostic for separating disk stars from halo substructures. These results are shown in Figure~\ref{fig:orbital_params} with the stars divided into sub-groups of metallicity (left), age (middle), and total velocity (right).

The region defined by $\ecc < 0.35$ and diskness $> 0.75$ was based on the results of \citet{Sales-Silva2019}, which in turn use a revision of the Geneva-Copenhagen Survey \citep{Casagrande+11} to identify the region occupied by the thin disk stars. All but one (SPLUS J013838.21-274012.0) non-MP contaminants are located within this region. In terms of ages, the region outside the thin disk selection is dominated by old and very old stars, while the thin disk region is equally populated by Young, Int. Age and Old stars. Out of all the young stars with measured orbital parameters, only two show high excursions ($\zmax > 2$ kpc) above the Galactic plane (SPLUS J104147.89-171551.9, SPLUS J132638.38-135134.3). 

\begin{figure}
 \includegraphics[width=\linewidth]{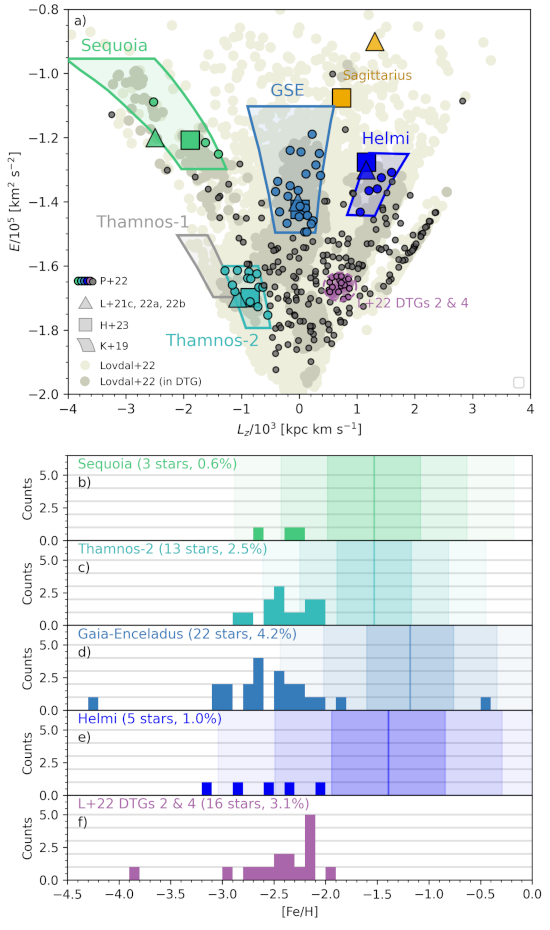}
 \caption{Top panel: $L_Z \times E$ for stars in the P+22 sample (circles with a black edge line), as compared to the regions occupied by several halo substructures from the literature: \citet[][K+19; shaded regions]{Koppelman+19}, \citet[][H+23; coloured squares]{Horta+2023}, \citet{Limberg+21c, Limberg+22, Limberg+22b} (L+21c, 22a, 22b; coloured triangles), as well as the DTGs from \citet[][Lovdal+22; violet and dark beige circles]{Lovdal+2022}. In the bottom panels, we show the metallicity distribution for the stars in the P+22 sample selected within each substructure. The vertical lines in the bottom panels are the metallicity values from the literature (H+23), whereas the shaded regions represent the 1, 2 and 3-$\sigma$ intervals.}
 \label{fig:substructures}
\end{figure}

For the velocities, we see that the three non-MP MeVeSt observed in Figure~\ref{fig:ages_feh_kin} are classified as thin disk stars following these dynamical criteria, while the only VMP LoVeSt is not. Therefore, for the purpose of identifying and eliminating the non-MP contaminants in the sample, a restriction in the $\ecc$-diskness diagram appears to provide better results than a selection using the Toomre diagram.

\subsubsection{Halo substructures}
\label{sec:substructures}

The Galactic halo is a mixture of stars formed in situ and the remnants of past mergers between the Milky Way and dwarf galaxies \citep{Springel+2005, Helmi+2020}. These remnants appear in the form of halo substructures and stellar streams, usually identified as stars that share similar dynamical properties \citep{Helmi+2000}. Each of these substructures has its own metallicity distribution function (MDF), which is related to the nature of the original dwarf galaxy and the merger event that originated the substructure \citep{Horta+2023}

In this Section, we analyse if the presence of possible members of these substructures in our dataset is biasing our targeting metallicity selection. For this analysis, we compare the total energy and vertical angular momentum in our sample ($E$ and $L_z$), to those expected for different halo substructures, streams and other dynamically tagged groups (DTGs) from the literature.

In the top panel of Figure \ref{fig:substructures} we plot the $E \times L_z$ parameters of the stars in the P+22 sample as grey circles (when not assigned to a substructure) and coloured circles (when assigned to a substructure). Five main Galactic substructures are represented by shaded regions defined as in \citep{Koppelman+19}: Sequoia \citep{Myeong+2019}; Gaia-Sausage/Enceladus \citep[GSE][]{Belokurov+2018, Helmi+2018}; Thamnos \citep{Koppelman+19}; and the Helmi streams \citep{Helmi+1999}. We note that Sequoia has also been suggested to be part of GSE \citep{Amarante+2022}. The stars in the P+22 sample within these regions are coloured accordingly. The coloured squares represent these substructures with $E$ and $L_z$ estimated by \citep[][hereafter H+23]{Horta+2023}, while the triangles represent the values obtained in a series of papers by \citet{Limberg+21c, Limberg+22, Limberg+22b}. The stars in the P+22 sample are distributed across the whole parameter space, sharing dynamical properties with several of these substructures (with the exception of Thamnos-1 and Sagittarius, where none or very few stars are present). They are also distributed both in regions assigned and not assigned to DTGs according to \citet{Lovdal+2022}.

To analyse the effects of possible substructure memberships in the selection of metal-poor stars, we compare the distribution of metallicities of the stars within the shaded regions with the metallicities estimated for each structure by H+23. These are shown in the lower panels of Figure \ref{fig:substructures}. The numbers in parenthesis show the number of stars selected in each substructure and the corresponding fraction of stars with respect to our whole sample. The vertical lines represent the H+23 metallicities and the shaded regions the 1, 2 and 3-$\sigma$ intervals around these values. As we can see, there is not a single star with measured metallicity within 1-$\sigma$ for any of the considered substructures. We also observe that in all cases, the stars are significantly more metal-poor (except for one star in Gaia-Enceladus) than reported in the literature for the substructures.

We also include in the top panel of Figure~\ref{fig:substructures}, as shaded beige circles in the background, the sample of \citet{Lovdal+2022}, with $E$ and $L_z$ recalculated by us using astrometric data from Gaia DR3. These authors assign the stars to different DTGs based on a hierarchical clustering method in 3-dimensional integrals of motion space. The stars assigned to a DTG are represented as darker beige circles. We highlight the DTGs 2 and 4 of \citet{Lovdal+2022} as violet circles. There is a concentration of stars in the P+22 sample around these groups, also represented in red. A 3-sigma selection around the centroid of these DTGs is represented by the dashed red ellipse, which contains 16 stars from the P+22 sample. We show the metallicity distribution of these stars in the bottom panel of Figure \ref{fig:substructures} (there is currently no known MDF for these groups in the literature). For the stars in the P+22 sample, the MDF for \citet{Lovdal+2022} DTGs 2+4 peaks at around [Fe/H] $= -2.1$, has an average of $-2.4$ and standard deviation of $0.4$. A star with metallicity $-3.82$ is also present within these groups. However, as evidenced by the MDF of the other known substructures, we point out that our selection is significantly biased towards VMP stars, which is also likely to affect the MDF observed for the DTGs. At best, our sample is representative of the metal-poor tail of the MDF of these DTGs.

From this analysis, we can conclude the methodology employed in the selection of metal-poor stars in S-PLUS is successful even among the stars that share the kinematical properties of different halo substructures, known to be, on average, more metal-rich than the stars we are targeting. Conversely, we see that it is also possible to use S-PLUS to target the metal-poor end of the MDF of these substructures.

\begin{figure*}
 \includegraphics[width=\linewidth]{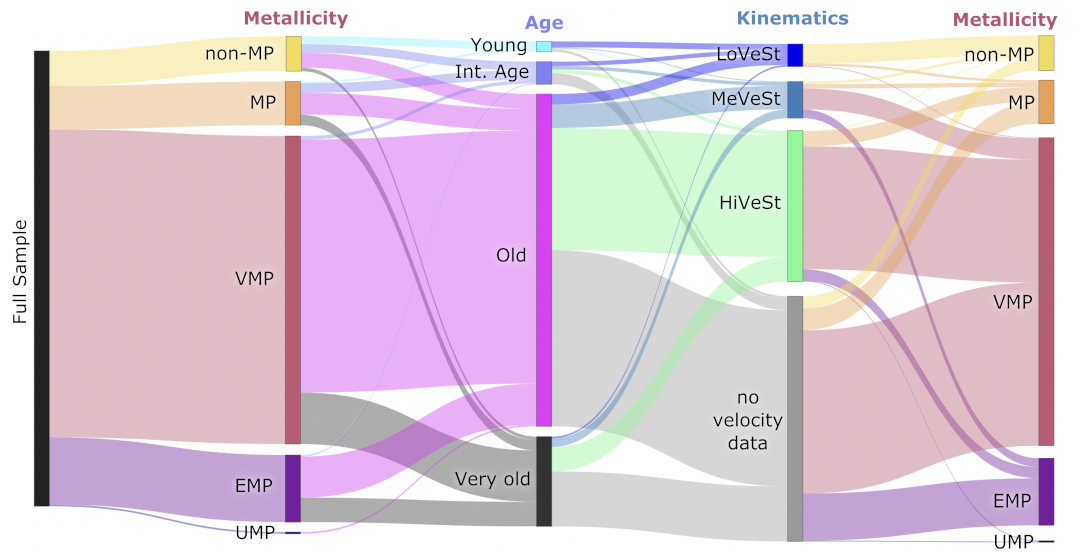}
 \caption{Sankey diagram presenting the distribution of stars among the sub-groups in metallicity, age and $V_\mathrm{Tot}$. The stars are divided into a) 5 metallicity groups: non-MP ([Fe/H] $> -1$), MP ($-2 <$ [Fe/H] $\leq -1$), VMP ($-3 <$ [Fe/H] $\leq -2$), EMP ($-4 <$ [Fe/H] $\leq -3$), UMP ([Fe/H] $\leq -4$); b) 4 age groups: Young ($\tau_\mathrm{adop} \leq 4$ Gyr), Int. Age ($4 < \tau_\mathrm{adop} \leq 7$ Gyr), Old ($7 < \tau_\mathrm{adop} \leq 10$ Gyr), Very Old ($\tau_\mathrm{adop} > 10$ Gyr); c) 4 $V_\mathrm{Tot}$ groups: LoVeSt ($V_\mathrm{Tot} \leq 80$ km s$^{-1}$); LoVeSt ($80$ km s$^{-1}$ $< V_\mathrm{Tot} \leq 180$ km s$^{-1}$); HiVeSt ($V_\mathrm{Tot} > 180$ km s$^{-1}$), and stars with no velocity data currently available.}
 \label{fig:sankey}
\end{figure*}

\subsection{Summary of group properties}

The Sankey diagram, Figure~\ref{fig:sankey}, is a useful tool to simultaneously visualize the distribution of classes for each parameter (metallicity, age and total velocity), as well as the correlations between each of their subgroups. Our results can be summarized as follows:

i) The sample is dominated by old and very old VMP stars. The second most numerous metallicity group are the EMP stars, where the fraction of very old stars is greater than any other group.

ii) Most of the young stars are contaminant non-MP stars. However, the age distribution in this non-MP group is similar to the overall distribution in the sample. For the MP group, the fraction of old and very old stars already starts to increase, but a significant number of Intermediate-age stars can also be found within this group.

iii) Roughly half of the sample does not have 6D astrometric information available. However, this limitation does not appear to bias the results of this work, as both the age and metallicity distributions within this group are similar to those in the full sample.

iv) The younger age groups have a larger fraction of LoVeSt, while older groups have increasingly larger fractions of HiVeSt. This is mostly a result of the disk population, composed mostly by the LoVeSt, covering a larger age range, including the youngest stars, while the MeVeSt and HiVeSt can be associated with the thick disk/halo populations, which have narrower and older age distribution. 

v) For the metallicity, the LoVeSt stars dominate the non-MP contaminants and contain only a very small fraction of MP and VMP stars. Not a single EMP or UMP star in this sample belongs to the LoVeSt group. 

vi) The fraction of non-MP contaminants decreases significantly for the MeVeSt group, and even more for the HiVeSt, which has only a single non-MP star. As mentioned before, this result indicates that the total velocity, and quantities derived from it, can be a powerful tool to aid in the selection of metal-poor stars for spectroscopic follow-up.

\subsection{Notable outliers} \label{sec:outliers}

In this Section we discuss the five notable outliers present in our sample. Their parameters are summarized in Table \ref{tab:outliers}.

\begin{table*}
\caption{Atmospheric parameters (estimated by P+22), mass and age (estimated in this work), astrometric parameters (from Gaia DR3) and orbital parameters (estimated in this work) for the five notable outliers discussed in Section \ref{sec:outliers}.}
\label{tab:outliers}
\begin{tabular}{llrrrrr} \hline \hline
                     & & \multicolumn{5}{c}{S-PLUS ID}                                                                                \\
                     & & J110831.29-223514.5 & J013838.21-274012.0 & J132638.38-135134.3 & J104147.89-171551.9 & J210428.01-004934.2  \\ \hline
$T_{\mathrm{eff}}$   & [K]             & $5234 \pm 100$   & $4980 \pm 100$   & $6524 \pm 100$   & $4251 \pm 100$   & $5056 \pm 100$   \\
{[}Fe/H{]}           & dex             & $-2.39 \pm 0.20$ & $-0.44 \pm 0.20$ & $-1.30 \pm 0.20$ & $-3.13 \pm 0.20$ & $-4.29 \pm 0.20$ \\
$\log g$             & dex             & $2.64 \pm 0.35$  & $3.59 \pm 0.35$  & $2.86 \pm 0.35$  & $0.57 \pm 0.35$  & $3.18 \pm 0.35$  \\ \hline
$m_{\mathrm{init}}$  & [$M_\odot$]     & $0.865$          & $0.894$          & $1.14$           & $1.92$           & $0.51$           \\
$\tau_\mathrm{ML}$   & [Gyr]           & $10.91$          & $14.51$          & $1.71$           & $0.51$           & $14.51$          \\
$\tau_\mathrm{E}$    & [Gyr]           & $ 9.54$          & $10.01$          & $4.35$           & $0.96$           & $8.08$           \\
$\tau_{50}$          & [Gyr]           & $ 9.48$          & $10.30$          & $3.14$           & $0.58$           & $8.14$           \\
$\tau_{16}$          & [Gyr]           & $ 5.88$          & $6.40$           & $1.40$           & $0.25$           & $2.97$           \\
$\tau_{84}$          & [Gyr]           & $12.84$          & $13.20$          & $7.06$           & $1.26$           & $12.61$          \\ \hline
$\alpha$ (J2000)     & degree          & $167.13034$      & $24.65924$       & $201.65991$      & $160.44952$      & $316.11672$      \\
$\delta$ (J2000)     & degree          & $-22.58734$      & $-27.67002$      & $-13.85954$      & $-17.26441$      & $-0.82621$       \\
$\mu_\alpha \cdot \cos{\delta}$ & mas yr$^{-1}$ & $2.508$ & $10.229$         & $-8.828$         & $-0.870$         & $14.976$         \\
$\mu_\delta$         & mas yr$^{-1}$   & $-2.133$         & $-3.297$         & $-0.285$         & $-0.359$         & $-8.260$         \\
$V_\mathrm{LOS}$     & \kms            & $20.01$          & $207.6$          & $275.56$         & $100.22$         & $-108.47$        \\
distance             & kpc             & $3.153$          & $5.554$          & $4.772$          & $9.688$          & $4.797$          \\ \hline
U                    & \kms            & $ 49.46$         & $-183.28$        & $-23.77$         & $-33.46$         & $-200.13$        \\
V                    & \kms            & $-19.22$         & $-256.08$        & $-251.43$        & $-99.92$         & $-205.52$        \\
W                    & \kms            & $  2.55$         & $-155.04$        & $228.18$         & $28.43$          & $-284.08$        \\
\Rapo                & kpc             & $12.25$          & $18.07$          & $10.59$          & $14.08$          & $20.42$          \\
\Rperi               & kpc             & $ 5.12$          & $0.08$           & $4.29$           & $5.68$           & $2.43$           \\
\zmax                & kpc             & $ 2.22$          & $13.34$          & $10.59$          & $7.76$           & $18.76$          \\
\ecc                 &                 & $ 0.41$          & $0.99$           & $0.42$           & $0.42$           & $0.79$           \\
diskness             &                 & $ 0.69$          & $0.15$           & $0.00$           & $0.29$           & $0.04$           \\
$E / 10^5$           & km$^2$ s$^{-2}$ & $ -1.519$        & $-1.414$         & $-1.586$          & $-1.450$        & $-1.34$          \\
$L_Z / 10^3$         & kpc km s$^{-1}$ & $ 1.659$         & $0.012$          & $-0.009$          & $1.488$         & $-0.377$         \\ \hline \hline
\end{tabular}
\end{table*}

\subsubsection{SPLUS J110831.29-223514.5: a VMP LoVeSt}

With $(U, V, W) = (49.5, -19.2, 2.6) \, \mathrm{km} \, \mathrm{s}^{-1}$, this star is classified as a LoVeSt. This is the only VMP star in this velocity group. However, when analysing dynamical parameters, this star does not share the same orbital properties as the other disk stars (in terms of $\ecc$, $z_\mathrm{max}$ and diskness). It is located in the same region in the age-metallicity diagram as MeVeSt, which shares similar dynamical properties. Considering its orbital parameters, we conlcude that it is in fact a thick disk star that just happens to be in a slower region in its trajectory. 

\subsubsection{SPLUS J013838.21-274012.0: a non-MP HiVeSt}

Opposite to the previous case, this star is the only HiVeSt classified as non-MP in this sample. This star also has a very high eccentricity and is one of our oldest non-MP stars. One possible scenario that could explain the existence of this star is the mechanism of disk heating \citep{Purcell+2010, Amarante+2020}: an initially non-MP LoVeSt formed in the disk would have its trajectory perturbed by larger mass structures in the Galaxy, causing a deviation from its originally circular orbit, increasing its UVW velocities. The very old age and high eccentric orbit of this star corroborate this hypothesis.

Another hypothesis to explain the nature of this star would be that it was accreted in the Galactic halo in a past merger event. We note that this star shares the dynamical properties of the Gaia-Enceladus substructure and has a metallicity within $2\sigma$ of the metallicity range attributed to this substructure (see Section \ref{sec:substructures}). However, accreted stars are expected to be more metal-poor than stars formed in the Milky Way at the same age, thus making this accretion hypothesis less likely.

\subsubsection{SPLUS J132638.38-135134.3: a Young HiVeSt}

This is the youngest HiVeSt in the sample and one of four HiVeSt with an age of less than 6 Gyr. These four stars are located in the region of the Kiel diagram where the turn-off of younger isochrones intercepts the horizontal branch of older isochrones. In this case, it is more likely that these stars are outliers with respect to stellar ages (when compared to other stars of the same velocity group) due to this degeneracy between younger/older isochrones other than to true physical property.

\subsubsection{SPLUS J104147.89-171551.9: a Young EMP star}

This star is the most significant outlier in the age-metallicity diagram shown in Figure~\ref{fig:ages_feh_kin} by being the only young star with [Fe/H] $< -2$. In this case, we also believe that the cause is a mistakenly attributed isochronal age due to mass-age degeneracy. This star is the coldest and has the lowest surface gravity in the sample and is located in a region in the Kiel diagram populated by a significantly higher number of high-mass and very-young stars. While the isochronal method returned a mass lower than 1.2 $M_\odot$ for more than 99\% of the sample, this star was assigned a mass of 1.9 $M_\odot$, explaining the considerably underestimated value characterized for its age. We also note EMP stars are expected to be low-mass stars given that they are likely exquisitely old.

\subsubsection{SPLUS J210428.01-004934.2: a peculiar UMP star}

This UMP star has the same dynamical properties (in terms of $E$ and $L_z$) of the GSE substructure and is already known for its peculiar carbon abundance \citet{Placco+21}, which is much lower than expected for this metallicity regime.

The surprising characteristic in this study is the apparently young age of SPLUS J210428.01-004934.2 in comparison to other HiVeSt. This can be explained by the fact this star belongs to the 6\% of our sample that had the isochronal method return a very flat age pdf, with the difference between the median and most-likely age being higher than 6 Gyr. This means that more UMP stars need to be observed for us to be able to discuss any meaningful insights about their expected ages.

\subsection{Improving the selection of metal-poor stars}

\label{sec:selection_hists}

\begin{figure*}
 \includegraphics[width=\linewidth]{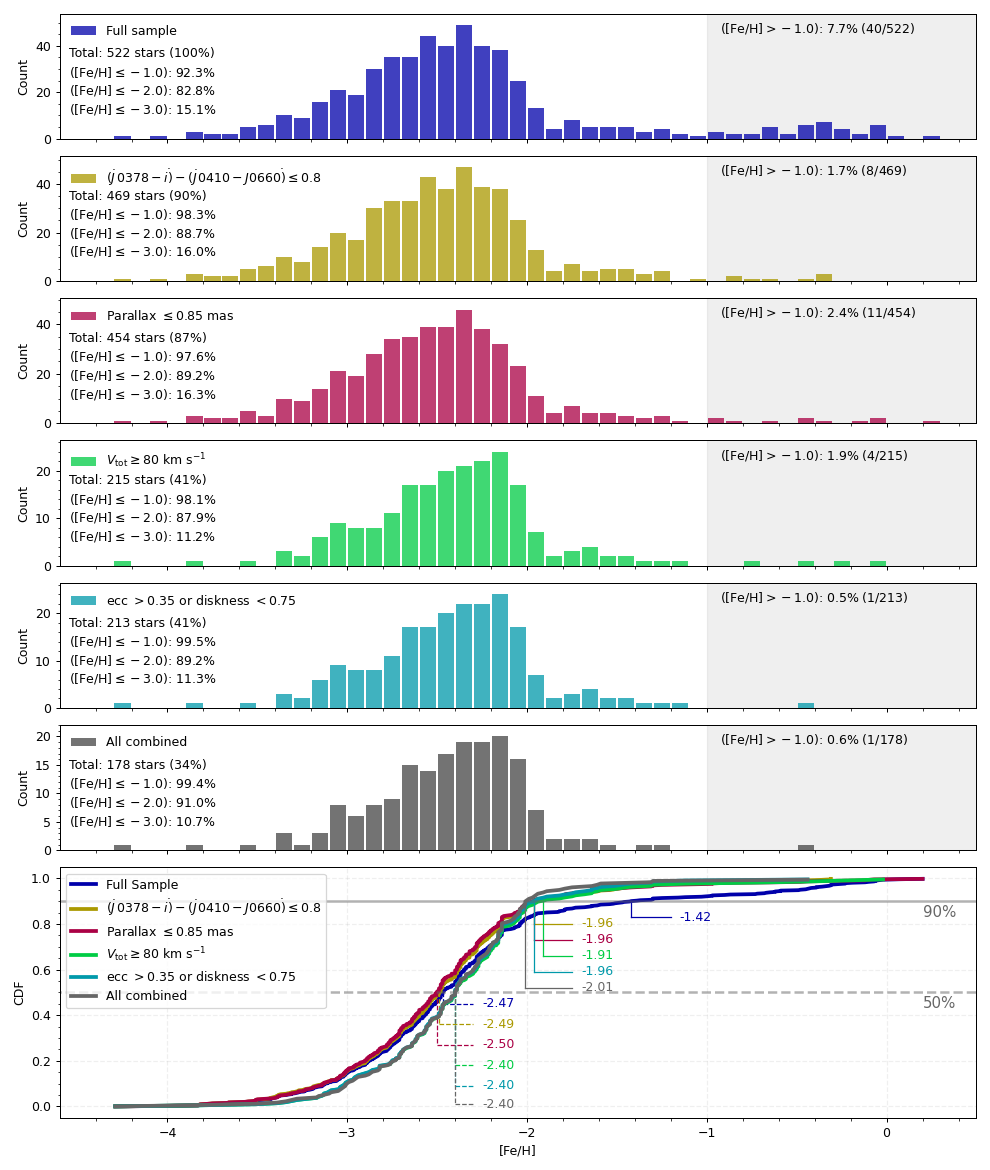}
 \caption{Metallicity distributions for the original sample of P+22 (top) and after applying the 5 different proposed selection criteria (middle panels). For comparison, the cumulative density functions for the six distributions are shown in the bottom panel, with the numbers indicating the values of the 50\% and 90\% percentiles.}
 \label{fig:feh_hists}
\end{figure*}

Finally, we apply our results to propose additional selection criteria that can help improve the accuracy of selecting metal-poor star candidates. These cuts are to be applied after the initial selection in the colour-colour diagram proposed by \citet{Placco+22}: $(J0395-J0410)-(J0660-J0861) \in [-0.30:0.15]$ and $(J0395-J0660)-2\times(g-i) \in [-0.60:-0.15]$. Here we consider 5 different additional criteria (each applied independently, except for the last one): 

i) Photometric ($[(J0378 - i)-(J0430 - J0660)] \leq 0.8 $); 

ii) Astrometric (Parallax $\leq 0.85$ mas); 

iii) Kinematical ($V_{Tot} \geq 80 \, km \, s^{-1}$); 

iv) Dynamical ($\ecc > 0.35$ or diskness $< 0.75$); 

v) All combined.

We note that the effectiveness of the astrometric and kinematical cuts is strongly dependent on the nature of our sample, which contains mostly high Galactic latitude stars (99\% of the stars in the sample have $|b| > 15 \deg$). Therefore, the selection of distant stars also implies the selection of stars with a higher distance from the Galactic disk plane, which in turn removes the non-MP thin disk contaminants from the sample. We expect the dynamical cut to be invariant to the Galactic latitude as it takes into account not only the current position, but the whole Galactic orbit.

The selection results are shown in Figure~\ref{fig:feh_hists}. The top panel presents the metallicity distribution in the original P+22 sample, while each of the five subsequent panels shows the distribution after applying the additional indicated restriction. Each of these panels also indicates the number of stars remaining in the sample after the selection and the fraction within each metallicity range. The amount of non-MP contaminants is indicated on the right. The bottom panel presents the cumulative density function (CDF) for each distribution, highlighting the 50\% and 90\% percentiles.

The dynamical selection is the one resulting in the highest purity of metal-poor stars (99.5\%), followed by the kinematical selection (98.5\%). However, both selections are limited to stars with 6D astrometric data, resulting in a cut of half the sample due to unavailable data. This limitation also removes several VMP and EMP from the sample, causing the median metallicity to be even higher than the original sample. A disadvantage of the dynamical cut is the necessity of adopting a Galactic potential model for orbit integration.

The photometric and astrometric selections result in nearly identical distributions, even though these parameters are not directly related. The photometric cut is slightly better than the parallax cut in terms of purity of metal-poor stars (98.3\% against 97.6\%), while the parallax cut gives slightly better results for the fraction of stars with $\metal \leq -3$ (16.3\% against 16.0\%). These differences are negligible given the size of the sample, and both selections can be used interchangeably. It is worth pointing out that this photometric selection involves the J0378 filter, which is known to have the largest calibration uncertainty in S-PLUS \citep[$\sigma_\mathrm{ZP} \sim 25$ mmags][]{Almeida-Fernandes+22}, whereas the astrometric selection has the disadvantage of being highly sensitive to the spatial distribution of the current sample and is expected to become less effective as S-PLUS covers regions of lower galactic latitude because of the increasing contamination of disk stars.

The combined cut is the one that presents the best result in terms of selecting the VMP stars (91\%, against 82.8\% from the original sample). However, it carries on the removal of the EMP stars from the kinematical and dynamical cut, resulting in 10.7\% of the sample being EMP (compared to 15.1\% in the original sample).

Given these considerations, we suggest the use of the photometric cut to significantly improve the purity of the metal-poor selection, optimizing the target selection for the spectroscopic follow-ups.

\section{Conclusions} \label{sec:conclusions}

\citet{Placco+22} used S-PLUS to select metal-poor candidates and showed through medium-resolution spectroscopic follow-up that 92\% of the 522 selected stars indeed have [Fe/H] below $-1$, while 83\% and 15\% have metallicities below $-2$ and $-3$, respectively.

In this work, we characterized the stars in this sample in terms of age and kinematical properties and analysed the correlations between these quantities and the metallicities in order to propose ways to further optimize the selection of metal-poor stars. The ages were calculated using a bayesian isochronal method, similar to the one presented by \citet{Jorgensen+Lindegren05}. Kinematical properties were derived from the 6D astrometric parameters from Gaia's DR3 and distances from \citet{Bailer-Jones+2021}. Orbital parameters were obtained after integrating the Galactic orbits for 10 Gyr for a \citet{McMillan2017} Galactic potential using \texttt{Galpy}.

We find the majority of the non-metal-poor contaminants in the sample are low-velocity stars, with ages distributed between ~2 and ~10 Gyr, which can be associated with the thin disk. The metal-poor stars, specially those with metallicity below $-2$, constitute the majority of the sample (83\%) and are medium/high velocity stars ($V_\mathrm{tot} > 80 \, km \, s^{-1}$). We also show that the presence of different substructures in the halo does not seem to be biasing the selection of metal-poor stars through colour-colour cuts in the S-PLUS photometry.

We propose five distinct selection criteria to be applied in addition to the ones described in \citet{Placco+22}, which can further improve the selection of metal-poor candidates. The dynamical selection ($\ecc > 0.35$ or diskness $< 0.75$) is the one that gives the best results in terms of optimizing the purity of the sample, resulting in 99.5\% of the stars having spectroscopic metallicity below $-1$, however, it should be used with care, as it introduces a dynamical bias in the selected sample. The selection of very metal-poor stars is optimized in the combined selection (where 91\% of the stars have metallicity below $-2$). For targeting the extremely metal-poor stars, the photometric cut ($[(J0378 - i)-(J0430 - J0660)] \leq 0.8 $) is the best option, resulting in a sample of 16\% of the stars having metallicity below $-3$. 

Applying the same restrictions as P+22 (\texttt{CLASS\_STAR} $\geq 0.95$; $g \leq 17.5$; $0.2 \leq g - i \leq 1.6$; and $0.3 \leq J0410 - J0861 \leq 3.5$) and the photometric selection criteria to the forthcoming S-PLUS DR4 (in preparation) results in a sample of $26.187$ stars, in which more than 98\% can be expected to be metal-poor according to our results. If we increase the selection for stars with $g$ brighter than 20.5 (this filter photometric depth for a signal-to-noise $= 5$) the sample size increases to $112.991$. Since DR4 only covers a third of the planned S-PLUS footprint, we expect to be able to select a sample of ${\approx}300.000$ high-confidence metal-poor stars by the end of the survey. We expect the results from this paper to contribute to the interpretation of the statistical results that will be derived from the sample of metal-poor stars selected in S-PLUS.

\section*{Acknowledgements}

First of all, we would like to thank the reviewer, whose valuable comments contributed to increase the quality of this paper. We thank Aaron Dotter, for the help provided in obtaining the MIST isochrones data; and S. Daflon, C. E. Ferreira Lopes and Y. Martins-Franco for useful comments and insights. The work of V.M.P. is supported by NOIRLab, which is managed by the Association of Universities for Research in Astronomy (AURA) under a cooperative agreement with the National Science Foundation. F.A.-F. acknowledges funding for this work from FAPESP grants 2018/20977-2 and 2021/09468-1. G.L. acknowledges FAPESP (procs. 2021/10429-0
and 2022/07301-5). L.B.S acknowledges the support of NASA-ATP award 80NSSC20K0509 and U.S. National Science Foundation AAG grant AST-2009122. J.A. acknowledges funding from the European Research Council (ERC) under the European Union’s Horizon 2020 research and innovation programme (grant agreement No. 852839). H.D.P. thanks FAPESP (procs. 2018/21250-9 and 2022/04079-0). The S-PLUS project, including the T80-South robotic telescope and the S-PLUS scientific survey, was founded as a partnership between the Funda\c{c}\~{a}o de Amparo \`{a} Pesquisa do Estado de S\~{a}o Paulo (FAPESP), the Observat\'{o}rio Nacional (ON), the Federal University of Sergipe (UFS), and the Federal University of Santa Catarina (UFSC), with important financial and practical contributions from other collaborating institutes in Brazil, Chile (Universidad de La Serena), and Spain (Centro de Estudios de F\'{\i}sica del Cosmos de Arag\'{o}n, CEFCA). We further acknowledge financial support from the S\~ao Paulo Research Foundation (FAPESP), Funda\c{c}\~ao de Amparo \`a Pesquisa do Estado do RS (FAPERGS), the Brazilian National Research Council (CNPq), the Coordination for the Improvement of Higher Education Personnel (CAPES), the Carlos Chagas Filho Rio de Janeiro State Research Foundation (FAPERJ), and the Brazilian Innovation Agency (FINEP). The authors who are members of the S-PLUS collaboration are grateful for the contributions from CTIO staff in helping in the construction, commissioning and maintenance of the T80-South telescope and camera. We are also indebted to Rene Laporte and INPE, as well as Keith Taylor, for their important contributions to the project. From CEFCA, we particularly would like to thank Antonio Mar\'{i}n-Franch for his invaluable contributions in the early phases of the project, David Crist{\'o}bal-Hornillos and his team for their help with the installation of the data reduction package \textsc{jype} version 0.9.9, C\'{e}sar \'{I}\~{n}iguez for providing 2D measurements of the filter transmissions, and all other staff members for their support with various aspects of the project.

\section*{Data Availability}

The data underlying this article will be shared on reasonable request to the corresponding author.



\bibliographystyle{mnras}
\bibliography{main} 




\appendix

\section{Verification of the isochronal method}

\label{sec:verification_isochronal}
\begin{figure*}
 \includegraphics[width=\linewidth]{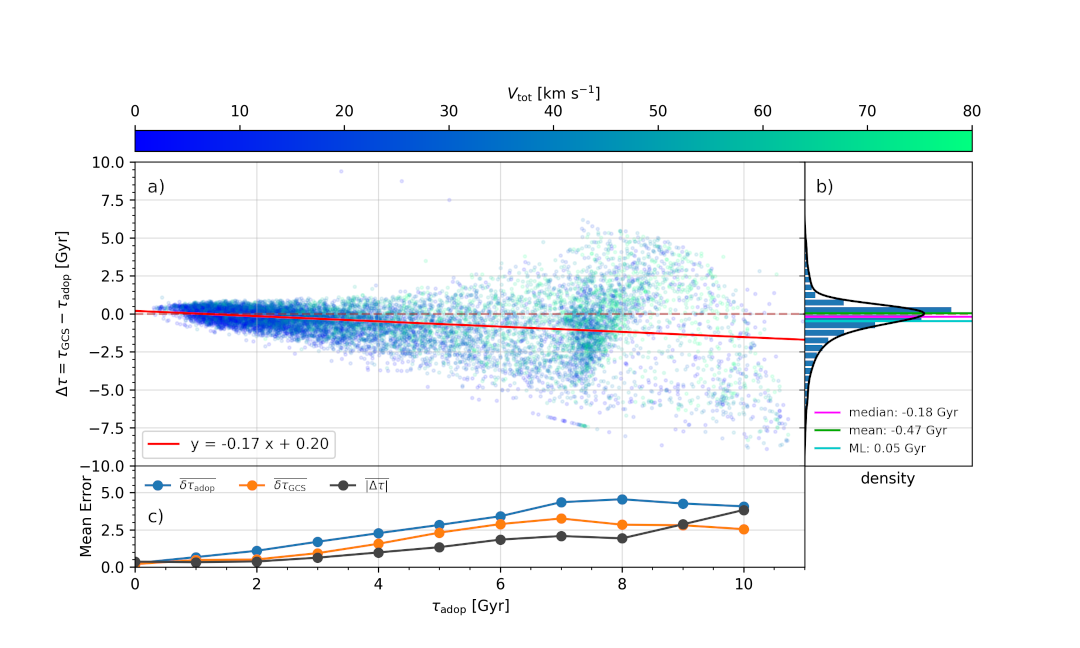}
 \caption{Panel \textit{a} shows the differences between the isochronal ages estimated by the method presented in this work and those estimated by \citet{Casagrande+11} for the Geneva-Copenhagen survey (all stars with an estimated age older than 0.0 Gyr), as a function of age. The stars are colour coded according to their velocity in the Local Standard of Rest. In Panel \textit{b} we represent the overall distribution of the age differences. Panel \textit{c} shows a comparison between the average differences and the average uncertainties in both methods in different age bins.}
 \label{fig:GCS_ages}
\end{figure*}

In this work, we presented a novel formalism for the implementation of the Bayesian isochronal age method (Section~\ref{sec:Methodology}). Our method is very similar to the one presented by \citet[][J\&L05]{Jorgensen+Lindegren05}. There are two differences in our implementation: i) our equations are written in a way that allows for the construction of a probability density function for any parameter predicted by the isochrones (e.g. age, initial mass, surface temperature, radius, luminosity); ii) we take into account the age-initial mass degeneracy by first obtaining a pdf for the initial mass from the observable inputs. Then, we adopt the median mass from this pdf as an additional input for the application of the isochronal method to derive the age pdf. 

Given these differences, it is good practice to verify our results against well-established stellar ages in the literature. We chose the Geneva-Copenhagen Survey \citep[GCS][]{Nordstrom+04} for this comparison, which contains data for 16682 nearby F and G dwarf stars. In particular, we use the data from \citet[][hereafter, C+11]{Casagrande+11}, which re-estimated the bayesian isochronal ages after improving the accuracy of the stellar effective temperatures. The differences between ours and C+11 approach are listed in Table~\ref{tab:gcs_differences}.

We estimated the ages of the GCS stars using the same method described in Section~\ref{subsec:ages}. In Figure~\ref{fig:GCS_ages} we compare our ages with the median ages derived by C+11 using the Padova isochrone set. The authors also provide ages characterized by different point estimators (mean and mode), and for another isochrone set (BaSTI). Our conclusions do not depend on the characterization chosen for the comparison.

In Panel \textit{a} of Figure~\ref{fig:GCS_ages}, we show the differences between our adopted ages and C+11 ages, as a function of the adopted ages. The scatter increases with stellar ages, but the offset remains relatively close to zero, with a small correlation shown by the linear regression (red line). For young stars, the offset is negligible and increases to $\sim 1.4$ Gyr for ages around 9.5 Gyr (which is the peak of the age distribution in the P+22 sample). The stars are coloured according to their total velocity in the Local Standard of Rest, and we observe an increase of higher velocity stars with age (a known property of the stars in the GCS). 

Panel \textit{b} shows the overall distribution of the age differences. The mode of the distribution is very close to zero (0.05 Gyr), but the median and mean are slightly higher (-0.18 Gyr and -0.47 Gyr, respectively). This is a result of the correlation between the offset and the ages.

In Panel \textit{c} we compare the average offset for different age bins (blue line) with the average uncertainties predicted by both methods for each respective bin. We define the uncertainty ($\delta$) as half the difference between the 84\% and 16\% percentiles of the age pdf (which for a gaussian distribution would approximately correspond to 1$\sigma$). Uncertainties calculated for the C+11 sample are represented by the orange line, while those calculated for the method presented in this work are shown in blue. The higher uncertainties in our case can be explained by the use of different inputs for the isochronal method and the use of more conservative uncertainties for the atmospheric parameters.

For the whole age interval, we observe that the average difference between ours and C+11 ages is smaller than the average estimated uncertainties. Therefore, we consider our ages reliable in comparison to those estimated for the GCS employing a similar bayesian approach.

\begin{table}
\caption{Summary of the differences in the implementation of the isochronal method between our approach and that of \citet{Casagrande+11} for the stars in the Geneva-Copenhagen survey.}
\label{tab:gcs_differences}
\begin{tabular}{rcc} \hline \hline
                  & This Work & \citet{Casagrande+11} \\ \hline
\textbf{Isochrone set}     & MIST      & Padova \footnote{\citet{Casagrande+11} also use BaSTI \citep{Pietrinferni+2004} isochrones. The comparison against both results in the same conclusions.}        \\
\textbf{Age prior}         & Uniform   & Uniform       \\
\textbf{Metallicity prior} & Uniform   & Uniform       \\
\textbf{IMF}               & Kroupa    & Salpeter      \\
\textbf{Inputs}            & \teff, \logg, \metal, \mass\footnote{The mass is determined from the resulting pdf of an initial run using the other parameters as the input}  & \teff, \metal, V               \\
\textbf{Point estimation}  & median    & median        \\ \hline \hline
\end{tabular}
\end{table}


\bsp	
\label{lastpage}
\end{document}